\documentclass[aps,prfluids,twocolumn,reprint,superscriptaddress,floatfix,longbibliography]{revtex4-1}
\usepackage{newtxtext,newtxmath}
\usepackage[T1]{fontenc}
\usepackage{ae,aecompl}
\usepackage{graphicx,color}
\usepackage{amsmath,amsfonts,amssymb,cancel} 
\usepackage{times}
\usepackage{hyperref}
\definecolor{NavyBlue}{rgb}{0.0,0,0.5}
\definecolor{Burgundy}{rgb}{0.5,0.0,0.125}
\hypersetup{
    breaklinks=true,
    colorlinks=true,
    citecolor={Burgundy},
    linkcolor={NavyBlue},
    urlcolor={NavyBlue}
}
\def\apj{Astrophys.\ J.}

\def\mnras{Mon.\ Not.\ R.\ Astron.\ Soc.}
\def\aap{Astron.\ Astrophys.}

\def\apjl{Astrophys.\ J.\ Lett.}

\def\jcap{J.\ Cosmol.\ Astropart.\ Phys.}
\def\physrep{Phys.\ Rep.}
\def\pre{Phys.\ Rev.\ E}
\def\prl{Phys.\ Rev.\ Lett.}

\def\araa{Ann.\ Rev.\ Astron.\ Astrophys.}

\newcommand{\Pm}{\text{Pr}_\mathrm{M}}
\renewcommand{\Re}{\text{Re}}          
\newcommand{\Rm}{\text{Re}_\mathrm{M}} 
\newcommand{\Rmc}{\text{Re}_\mathrm{M}^{\text{(crit)}}}    

\renewcommand{\vec}[1]{\mathbf{#1}}	
\newcommand{\dd}{\mathrm{d}}        
\newcommand{\ii}{\mathrm{i}}        

\newcommand{\kpc}{\,{\rm kpc}}  

\newcommand{\yr}{\,{\rm yr}}    

\newcommand{\brms}{\,b_{\rm rms}}
\newcommand{\urms}{\,u_{\rm rms}}

\newcommand{\lb}{l_b}
\newcommand{\lu}{l_u} 
\newcommand{\aub}{\cos(\theta)_{\vec{u}, \, \vec{b}}}
\newcommand{\ajb}{\cos(\theta)_{\vec{j}, \, \vec{b}}}
\newcommand{\aeob}{\cos(\theta)_{\vec{e}_1, \, \vec{b}}}
\newcommand{\aetwob}{\cos(\theta)_{\vec{e}_2, \, \vec{b}}}
\newcommand{\aetb}{\cos(\theta)_{\vec{e}_3, \, \vec{b}}}
\newcommand{\kf}{k_\mathrm{F}}
\newcommand{\Rmloc}{(\text{Re}_\mathrm{M})_\text{loc}}

\newcommand\Eq[1]{Eq.~(\ref{#1})} 
\newcommand\Fig[1]{Fig.~\ref{#1}}
\newcommand\Sec[1]{Section~\ref{#1}}
\newcommand\Tab[1]{Table~\ref{#1}}

\begin{document} 
\title{On the saturation mechanism of the fluctuation dynamo at $\Pm \ge 1$}
\author{Amit Seta}\thanks{amit.seta@anu.edu.au}
\affiliation{Research School of Astronomy and Astrophysics, Australian National University, Canberra, ACT 2611, Australia}
\affiliation{School of Mathematics, Statistics and Physics,
Newcastle University, Newcastle Upon Tyne, NE1 7RU, UK}
\author{Paul J. Bushby}
 \affiliation{School of Mathematics, Statistics and Physics,
Newcastle University, Newcastle Upon Tyne, NE1 7RU, UK}
\author{Anvar Shukurov}
 \affiliation{School of Mathematics, Statistics and Physics,
Newcastle University, Newcastle Upon Tyne, NE1 7RU, UK}
\author{Toby S. Wood}
 \affiliation{School of Mathematics, Statistics and Physics,
Newcastle University, Newcastle Upon Tyne, NE1 7RU, UK}
\date{\today}

\begin{abstract} 
The presence of magnetic fields in many
astrophysical objects is due to dynamo action, whereby a part of the kinetic energy
is converted into magnetic energy. A turbulent dynamo that produces magnetic field structures on the same scale as 
the turbulent flow is known as the fluctuation dynamo. We use numerical simulations to explore the 
nonlinear, statistically steady state of the fluctuation dynamo in driven turbulence.
We demonstrate that as the magnetic field growth saturates, its amplification and diffusion are both affected by the back-reaction of the Lorentz force upon the flow.
The amplification of the magnetic field is reduced due to stronger alignment between the velocity field,
magnetic field, and electric current density. Furthermore, we confirm that the amplification decreases due to a weaker stretching of the magnetic field lines.
The enhancement in diffusion relative to the field line stretching is quantified by a decrease in the computed local value of the magnetic Reynolds number. 
Using the Minkowski functionals, we quantify the shape of the magnetic structures produced by the dynamo as magnetic filaments and ribbons in both kinematic and saturated dynamos and derive the scalings
of the typical length, width, and thickness of the magnetic structures with the magnetic Reynolds number. We show that all three of these magnetic length scales increase as the dynamo saturates. 
The magnetic intermittency, strong in the kinematic dynamo (where the magnetic field strength grows exponentially) persists in the statistically steady state, but intense magnetic filaments and ribbons are more volume-filling.
\end{abstract}

\maketitle

\section{Introduction}
Magnetic fields are observed in a variety of astrophysical objects,
including
stars, galaxies and galaxy clusters, where they play an important 
role in various physical processes. Based on length and time scales, astrophysical magnetic fields can be divided into two types:
the large-scale or mean field, which is coherent over scales comparable
to the size of the system, and the small-scale or fluctuating field, 
whose correlation length is of the order of the driving scale of the underlying turbulent flow. 
The driving scale of turbulence, $l_0$, is of the order of $0.1 \kpc$ in spiral galaxies \citep{Gaenslar2005,Fletcher2011,Houde2013}, and $10 \kpc$ in galaxy clusters \citep{GF04,SC06}. 
The fluctuating magnetic field is believed to evolve over the eddy turnover timescale, which is considerably shorter than the corresponding evolution timescale for the large-scale field
(which is typically of the order of $10^{8} \yr$ in spiral galaxies, comparable to the rotation period).
For spiral galaxies, the mean and fluctuating fields have comparable magnitudes
and thus both kinds of fields are equally important
for the galactic dynamics \citep{Beck2016}. There are a number
of reviews covering the theoretical, numerical, and observational aspects
of the subject \citep{Beck1996,Widrow2002,
BS2005,KZ08,Fed16,Rincon19}. 

The evolution and maintenance of magnetic fields is generally explained by
dynamo action, a process by which 
kinetic energy is converted to magnetic energy.
Astrophysical flows leading to dynamo action are typically turbulent; such flows may be
driven by
convection in stars, supernovae
in galaxies, and merger shocks, motion of galaxies and AGN outflows in galaxy clusters.
Magnetic field amplification by turbulent motions has also been observed in laboratory experiments \citep{Tz18}.
Depending upon the magnetic fields that they produce, such dynamos are generally categorized as either
mean-field or fluctuation (or ``small-scale'') dynamos. 
Mean-field dynamos produce large-scale magnetic fields, whereas the
fluctuation dynamo
generates the small-scale component of the field via random stretching of field lines by the turbulent
velocity \citep{Kazantsev1968,ZRMS84} (as conceptually explained by the stretch-twist-fold mechanism \citep{CG95,SBS2015}). 
Fluctuation dynamo action plays a crucial role not only in spiral galaxies \citep{RSS88,KA92,Beck1996,SS2008,KZ08,Pakmor2017}, elliptical galaxies \citep{MS96,Seta19} and galaxy clusters \citep{RSS89,SSH06,BS13,Vazza2018}, 
but also in stars such as the Sun \citep{Cat1999,PCS10,Bushby14,Rempel14}, making it a general type of astrophysical process.
Fluctuation dynamos naturally produce
intermittent magnetic fields \citep{ZRS90,SCTMM04,Wilkin2007}, 
characterised by the presence of intense, localised field structures. 
In the galactic context, 
a better understanding of these structures is needed for cosmic ray propagation studies \citep{SSSBW17,SSWBS18} and in the galaxy cluster context
for the interpretation of radio observations \citep{EC06}. The initial stages of magnetic field growth, when the Lorentz force is negligible, have been thoroughly studied \citep{ZRS90,BS2005}, so here we focus on 
the nonlinear states of the fluctuation dynamo, for which it is possible to consider relatively simple idealised flows (i.e., homogeneous, isotropic turbulence). 
A mean-field dynamo would require additional physics, such as rotation, velocity shear and density stratification; such effects can be safely ignored over the length and time scales that will be of interest here.  

In a fluctuation dynamo, the root mean square (rms) magnetic field grows exponentially if the magnetic Reynolds number $\Rm$ (quantifying the efficiency of inductive effects compared to magnetic diffusion) exceeds its critical value $\Rmc$, which depends on the properties of the flow.
When the magnetic energy is low in comparison to the turbulent kinetic energy, the 
flow dynamics are not influenced by the 
magnetic field (the kinematic stage). For an isotropic, incompressible, mirror--symmetric, homogeneous and Gaussian random velocity field,
which is also $\delta$--correlated in time, 
it can be shown that the magnetic field power spectrum $M_k$ in the kinematic stage follows a power-law (at low wave numbers) with slope $3/2$ \citep{Kazantsev1968,BS2005}. 
However, an exponentially growing magnetic field also leads to the exponential growth of the Lorentz force, 
which eventually makes the problem nonlinear. This slows down the growth and finally leads to the saturation of the dynamo (the saturated stage).
The nonlinear problem is mostly studied via numerical simulations, in which the Navier-Stokes and induction equations are solved simultaneously 
\citep[e.g.,][]{Meneguzzi1981,Cat1999,Haugen2004,SCTMM04,CR09,CT09,Bushby14,Fed11,FB12,Sur12,Ber2012,BS13,Fed14,SBS18}. 
Our aim in this paper is to explore the saturation mechanism of the fluctuation dynamo and to characterize the magnetic structures it generates. 

For fluctuation dynamos driven by homogeneous and isotropic turbulence, the following three quantities are prescribed: the driving scale of the turbulent flow $l_0$, the fluid viscosity $\nu$, and the
magnetic resistivity $\eta$. Based on the magnetic Prandtl number $\Pm$ (defined to be the ratio of viscosity to resistivity, $\Pm = \nu/\eta$), 
fluctuation dynamos can be divided into small and large $\Pm$ cases.  
$\Pm$ is greater than unity ($\eta < \nu$) for hot diffuse plasma (interstellar and intergalactic medium) and $\Pm$ is much smaller than unity  ($\eta > \nu$)  for dense plasma
(planets, stars and liquid metal dynamo experiments). The critical magnetic Reynolds number $\Rmc$, which is a threshold for dynamo action to occur, increases
with decreasing $\Pm$ \citep{BC2004,Sch05,Sch07,Iskakov07,BHLS18}. We focus upon the 
$\Pm \ge 1$ regime, fixing the underlying flow (i.e., fixing $\Re$) and then varying $\Rm$ in order to study the sensitivity of the magnetic structures of nonlinear dynamo states to the magnetic Reynolds number.

This paper is structured as follows. In \Sec{sec:num}, we introduce the basic equations and describe the numerical setup and provide parameters of the simulations. In \Sec{sec:int}, we discuss magnetic field intermittency and in \Sec{sec:sat}, we examine
possible saturation mechanisms. Then, in \Sec{sec:morph}, we use Minkowski functionals to quantify the magnetic field structures (as a function of the magnetic Reynolds number) in both the kinematic and nonlinear regimes.
Finally, in \Sec{sec:dis}, we conclude with a discussion and propose some future directions of research.
 
\section{Basic equations and numerical modelling}
\label{sec:num}
To study the fluctuation dynamo action in a turbulent flow driven by a prescribed random force, 
we solve the equations of magnetohydrodynamics, using the Pencil code \footnote{Website: https://github.com/pencil-code}. 
The computational domain is a triply-periodic cubic box of non-dimensional width $L=2\pi$, with $256^3$ or $512^3$ grid points. The equations are solved with sixth-order finite differences in space and a third-order Runge--Kutta scheme for the temporal evolution. The governing equations are  
\begin{align}
\frac{\partial \rho}{\partial t} & + \nabla \cdot (\rho \vec{u})  = 0, \label{fdce} \\
\frac{\partial \vec{b}}{\partial t} & = \nabla \times (\vec{u} \times \vec{b}) + \eta \nabla^2 \vec{b},  \label{fdie} \\ 
\frac{\partial \vec{u}}{\partial t} & + (\vec{u} \cdot \nabla) \vec{u} = -\frac{\nabla p}{\rho} + \frac{ \vec{j} \times \vec{b}}{c\rho}  \nonumber \\
& + \nu \left(\nabla^2 \vec{u} + \frac{1}{3} \nabla (\nabla \cdot \vec{u}) + 2 \vec{S} \cdot \nabla \ln \rho \right) + \vec{F}, \label{fdns}
\end{align}
where $\vec{u}$ is the velocity field, $\vec{b}$ is the magnetic field, $\rho$ is the fluid density, $p$ is the pressure, $\eta$ is the magnetic diffusivity,
$\vec{j} = (c/4\pi)\nabla\times\vec{b}$ is the electric current density, 
$c$ is the speed of light, $\nu$ is the viscosity, $S_{ij} = \frac{1}{2} \left(u_{i,j} + u_{j,i} - \frac{2}{3} \delta_{ij} \nabla \cdot \vec{u} \right)$ is the rate-of-strain tensor, 
and $\vec{F}$ is the forcing function
(defined below). 
We use an isothermal equation of state, $ p =c_s^2 \rho $, where the constant $c_s$ is the sound speed.
\Eq{fdie} is solved in terms of the magnetic vector potential to ensure that the magnetic field remains divergence free.

We drive the flow with a mirror-symmetric and $\delta$-correlated in time forcing \citep{Haugen2004} of the form
\begin{align} 
\vec{F}(\vec{x},t) = {\rm Re}\{N \vec{F}_{\vec{k}(t)} \exp[\ii \vec{k}(t) \cdot \vec{x} + \ii \phi(t)]\},
\end{align}
where $\vec{k}$ is the wave vector, $\vec{x}$ is the position vector and $-\pi < \phi \le \pi$ is a random phase. 
To ensure that
the forcing is nearly $\delta$--correlated in time, $\vec{k}$ and $\phi$ are changed at each time step $\delta t$. Also, to ensure that the time-integrated force
is independent of the chosen time step $\delta t$, the normalization is $N = F_0 c_s (|\vec{k}| c_s/\delta t)^{1/2}$, where $F_0$ is the non--dimensional forcing amplitude chosen
such that the maximum Mach number is small enough ($\urms/c_s \lesssim 0.1$) to avoid strong compressibility.  
We select many random wave vectors $\vec{k}$, each of magnitude $k$ (a multiple of $2\pi/L$ to make sure that the flow is periodic) in a given range. 
Then we select an arbitrary unit vector $\vec{e}$ (neither parallel nor anti-parallel to $\vec{k}$) and set
\begin{align} \label{forcing}
\vec{F}_{\vec{k}} = \frac{\vec{k} \times \vec{e}}{|\vec{k} \times \vec{e}|}.
\end{align}
The form of \Eq{forcing} ensures that the 
forcing
is solenoidal, 
i.e. $\nabla \cdot \vec{F} = 0$ by construction.
The average wave number at which the flow is driven is denoted by $\kf$. Even when the flow is periodic, $\kf$ need not be a multiple of $2 \pi/L$.
Physically, $2 \pi/\kf$ represents the driving scale of the turbulent flow, $l_0$, in the system.

The turbulent plasma is characterized by the hydrodynamic Reynolds number $\Re$ and magnetic Reynolds number $\Rm$,
 defined in terms of the rms velocity $\urms$ and the forcing scale $\kf$\footnote{It is also common to define the hydrodynamic and magnetic Reynolds number with respect to the forcing wave number instead of the driving length scale.
Then the Reynolds numbers are smaller by a factor $2 \pi$ than the values we quote.}, as 
\begin{align}
\Re = \frac{\urms}{\nu} \frac{2 \pi}{\kf}, \, \,  \, \, \Rm = \frac{\urms }{\eta} \frac{2 \pi}{\kf}.
\label{rmdef}
\end{align}
We use non-dimensional units with lengths in units of the domain size $L=2\pi$, speed in units of the isothermal sound speed $c_s$,
time in units of the eddy turnover time $t_0 = 2 \pi /\urms \kf$, density in units of the initial density $\rho_0$ and the magnetic field in units of $\left(4 \pi \rho_0 c_{s}^2\right)^{1/2}$. Initially, the density is constant everywhere and $\vec{u}=\vec{0}$, whilst there is a weak random, seed magnetic field with zero net flux across the domain. 
 
\begin{table}
	\caption{Summary of fluctuation dynamo simulations in a numerical domain of size $(L = 2\pi)^3$ with $256^3$ mesh points.
        In all cases, the forcing scale $\kf$ is approximately equal to $1.5 (2 \pi/L)$, the forcing amplitude $F_0=0.02$, the magnetic Prandtl number $\Pm=1$ 
	and the rms velocity in the saturated state is $\urms/c_s \approx 0.11$. For each simulation, we quote 
	the Reynolds number, the magnetic Reynolds number,
	the rms magnetic field in the saturated state $\brms$, 
         the ratio of magnetic to kinetic energy in the saturated state $\varepsilon_{\rm M}/\varepsilon_{\rm K} = \brms^2/\urms^2$,
         the correlation length of the velocity and magnetic field in the kinematic stage ${\lu}_{\rm kin}$ and ${\lb}_{\rm kin}$, and similarly
         in the saturated stage ${\lu}_{\rm sat}$ and  ${\lb}_{\rm sat}$.}
	\label{table_nfd}
	\begin{tabular}{cccccccc} 
		\hline 
		$\eta,\nu$ & $\Rm, \Re$ & $\brms$  &  $\varepsilon_{\rm M}/\varepsilon_{\rm K}$ & ${\lu}_{\rm kin}$ & ${\lb}_{\rm kin}$ & ${\lu}_{\rm sat}$ & ${\lb}_{\rm sat}$\\
		\hline 
                 $10\times 10^{-4}$ & $449$ & $0.033$ & $0.08$ & $3.14$ & $1.82$ & $3.77$ & $1.95$\\
                 $5\times 10^{-4}$ & $898$ & $0.042$ & $0.14$ & $3.20$ & $1.26$ & $3.45$ & $1.76$\\
                 $4\times 10^{-4}$ & $1122$ & $0.048$ & $0.20$& $3.01$ & $0.94$ & $3.64$ & $1.76$\\
                 $3\times 10^{-4}$ & $1496$ & $0.049$ & $0.21$& $3.01$ & $0.88$ & $3.39$ & $1.57$\\
                 $2.5\times 10^{-4}$ & $1796$ & $0.054$ & $0.25$& $2.95$ & $0.75$ & $3.58$ & $1.57$\\
                 $2\times 10^{-4}$ & $2244$ & $0.055$ & $0.26$& $2.95$ & $0.69$ & $3.33$ & $1.56$\\
		\hline
	\end{tabular}
\end{table}

\begin{figure}
 {\includegraphics[width=\columnwidth]{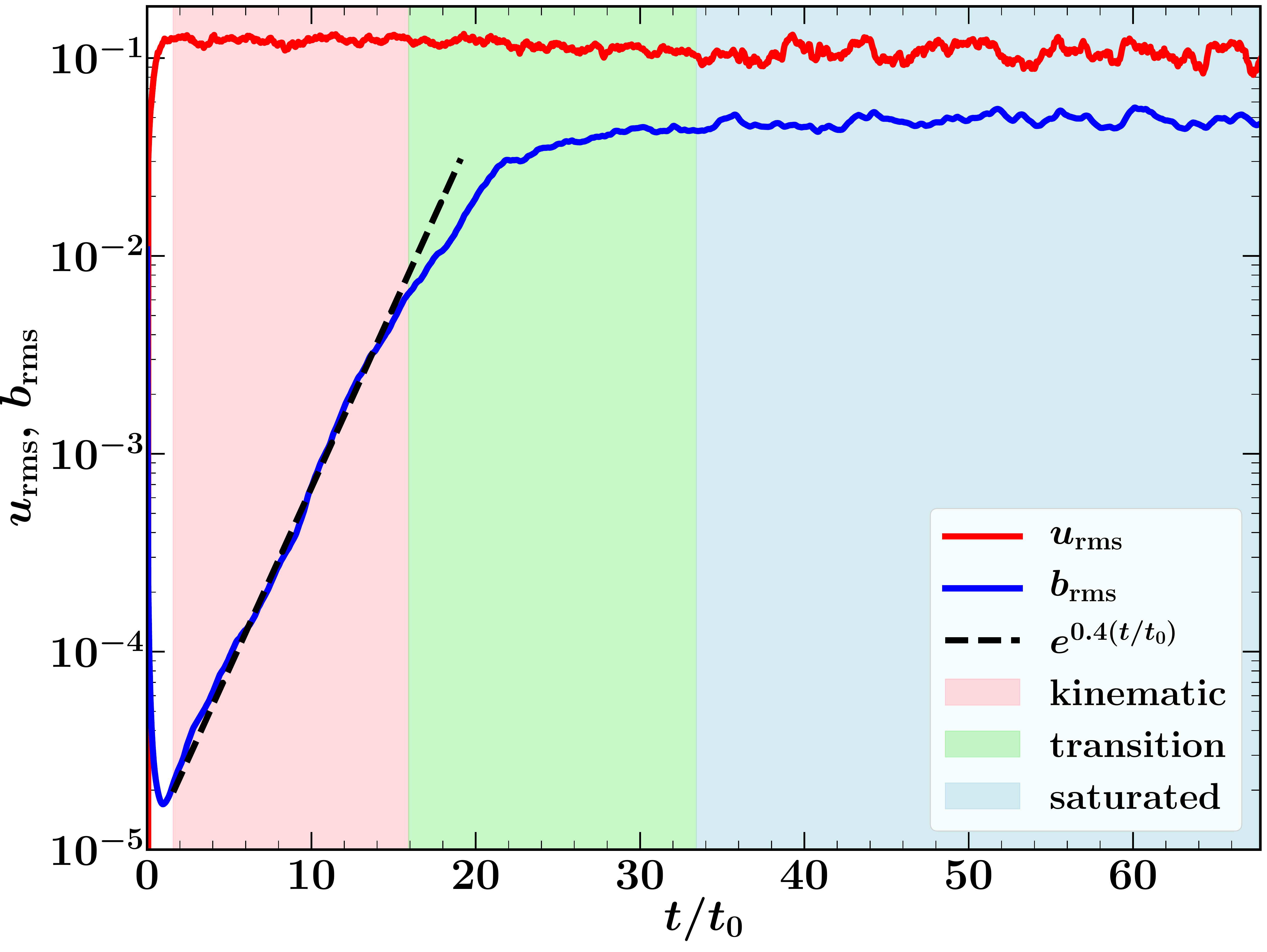}}
\caption{Root mean square (rms) velocity field $\urms$ (red) and magnetic 
field $\brms$ (blue) as functions of normalized time $t/t_0$ (where $t_0 = 2 \pi /\urms \kf$ is the eddy turnover time) 
for  
$\Re= \Rm = 1122$.
During the kinematic stage (area shaded in light red), the black dashed line corresponds to the exponential growth. 
As the magnetic field grows, the dynamo passes through a transitional stage (area shaded in light green), before reaching a statistically steady saturated state (area shaded in light blue).}
\label{ts} 
\end{figure}

For the first set of simulations, with parameters given in \Tab{table_nfd}, the turbulent motions are driven at the wave numbers $2 \pi/L$ and $2 (2 \pi/L)$ at equal intensities, which implies that $\kf \approx 1.5 (2 \pi/L)$. The magnetic field grows for $\Rm \ge \Rmc$, 
with $\Rmc \approx 220$ for $\Pm=1$ \citep{Haugen2004}. The evolution of the rms velocity field, $\urms$, and magnetic field, $\brms$, is shown in \Fig{ts} for $\Rm=1122$. The flow speed is controlled
by the forcing function and thus remains nearly constant. The magnetic field first decays until it  
reaches
an eigenstate of the induction equation.
Then it grows exponentially in the kinematic stage at the growth rate of $0.4 \urms\kf/2 \pi$ in dimensional units. As it becomes stronger, the Lorentz force affects the flow 
and slows down the exponential increase.  Finally, when the magnetic field becomes strong enough,
the dynamo
reaches a statistically steady state in the
saturated stage. The exponential growth and then saturation of the magnetic field occurs in all of the runs shown in \Tab{table_nfd}. 

\begin{figure}
 {\includegraphics[width=\columnwidth]{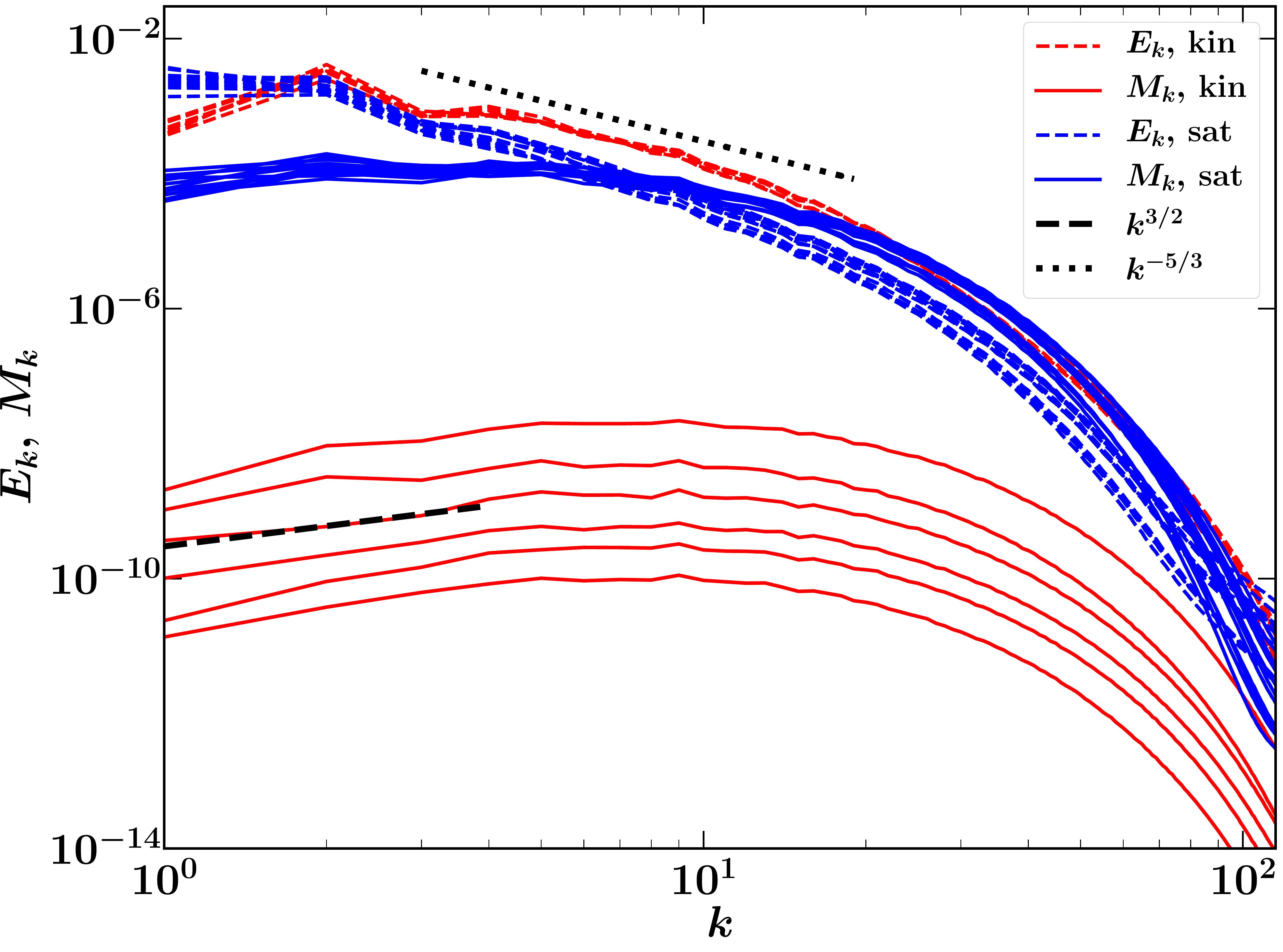}}
\caption{The shell--averaged (one--dimensional) kinetic $E_k$ (dashed) and magnetic $M_k$ (solid) energy spectra in the kinematic (red) and 
saturated (blue) stages for 
$\Re= \Rm = 1122$.
The kinetic energy spectrum is close to the Kolmogorov spectrum, $E_k \propto k^{-5/3}$ (dotted, black) in the
main part of the wave number range.
The magnetic spectrum is initially 
of the form
$M_k \propto k^{3/2}$ (dashed, black) at smaller wave numbers. As the magnetic field saturates, its power shifts to smaller wave numbers and the magnetic spectrum
flattens.}
\label{spec} 
\end{figure}

The shell-averaged (one-dimensional) power spectra, for various stages of the magnetic field evolution, are shown in \Fig{spec}. 
At all times, the kinetic energy spectrum is close to the Kolmogorov spectrum, $E_k \propto k^{-5/3}$, in the range $3 \le kL/2\pi \le 20$ (flow is driven at $k=2 \pi/L$ and $k=2 (2 \pi/L)$), 
which suggest that the velocity field is turbulent in nature.
The magnetic spectrum in the kinematic stage
has a broad maximum at large wave numbers and its slope agrees with the Kazantsev model, $M_k \propto k^{3/2}$, in the range $2 \le kL/2\pi \le 10$ with maximum power at approximately $kL/2 \pi=10$. 
Kazantsev's theory assumes that the turbulent flow is $\delta$-correlated in time.
Whilst we have used a $\delta$-correlated forcing in the Navier-Stokes equation (term $\vec{F}$ in \Eq{fdns}), the flow that it drives is not $\delta$-correlated, especially at high $\Re$.
However, it is known that the slope of the spectrum in the kinematic stage remains the same even when the flow has a finite but small correlation time \citep{BS14,BS15}, which explains why we recover the Kazantsev result in these simulations. 
As the magnetic field grows, the spectral maximum
shifts to smaller wave numbers and the spectrum becomes much flatter with a broad maximum in the range $2 \le kL/2\pi \le  5$.

\section{Magnetic intermittency} 
\label{sec:int}
Intermittency in a random field can manifest itself via heavy tails in its probability distribution function (PDF) and
leads to an increased kurtosis in comparison with the Gaussian distribution. For the random velocity field $\vec{u}$ with zero mean, the kurtosis is defined by
\begin{align}
{\rm K}(\vec{u}) = \frac{\langle \vec{u}^4 \rangle}{\langle \vec{u}^2 \rangle^2},
\label{kurtu}
\end{align}
with angular brackets denoting the volume average. A useful diagnostic of the spatial structure is the correlation length of the field, $\lu$, which is calculated from the power spectrum $E_k$ as
\begin{align}
\lu =  \frac{\pi}{2} \frac{\int_0^{\infty} 2 \pi k^{-1} E_k \, \dd k}{\int_0^{\infty} E_k \, \dd k}.
\label{lcu}
\end{align}
Here, using such tools, we discuss the spatial intermittency of the velocity
and magnetic fields in nonlinear fluctuation dynamos.

\begin{figure}
 {\includegraphics[width=\columnwidth]{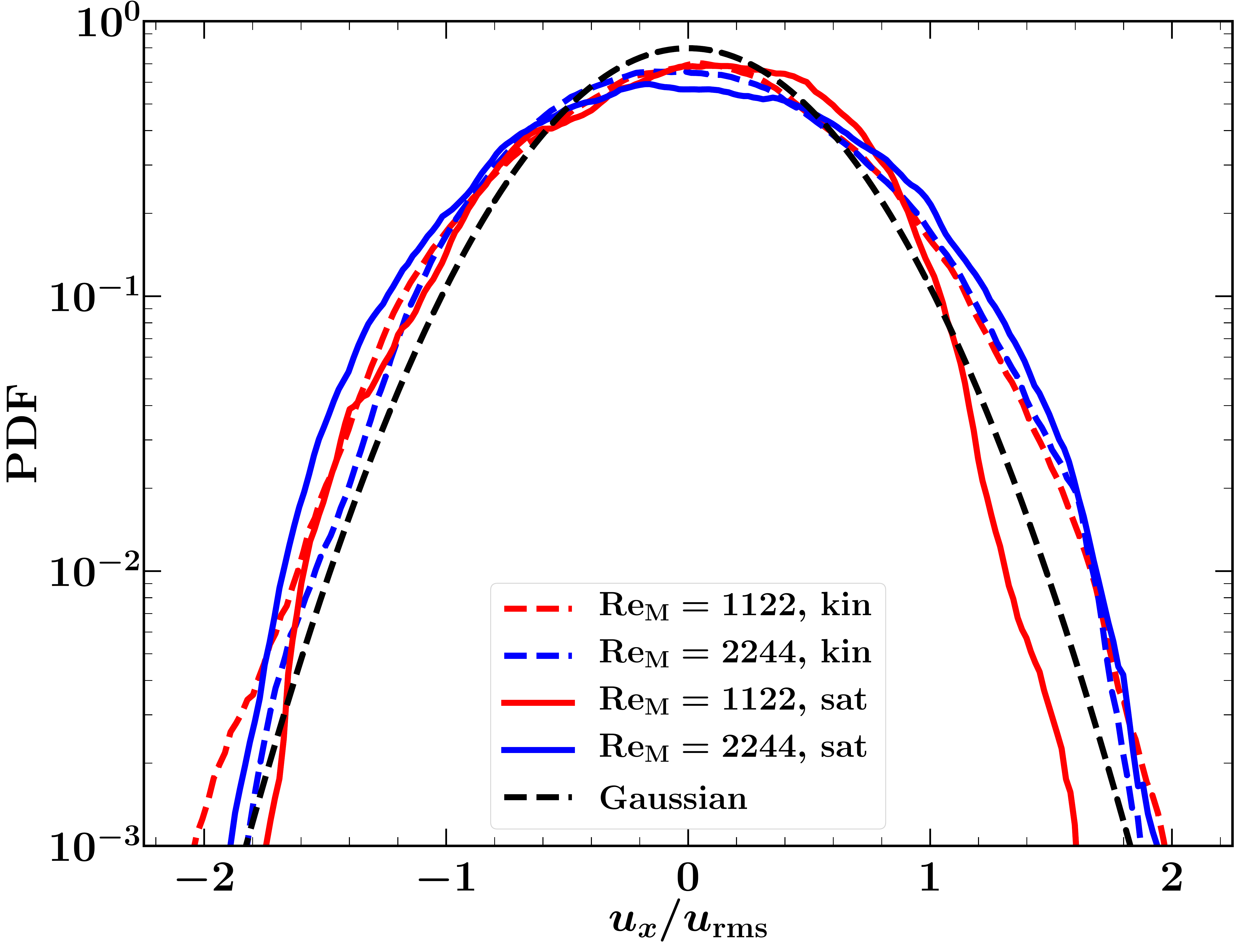}}
\caption{The PDF of the normalized velocity field component $u_x/\urms$ for $\Rm=1122$ and $\Rm=2244$
in the kinematic (dashed) and saturated (solid) stages for the value of $\Rm$ given in the legend. 
The PDF of the single component of the velocity field 
is roughly Gaussian (dashed, black) in both the stages for both $\Rm$. Here only $u_x/\urms$ is shown but similar behaviour is exhibited by all three velocity components.}
\label{updf} 
\end{figure}

\begin{figure*}
\includegraphics[width=\columnwidth]{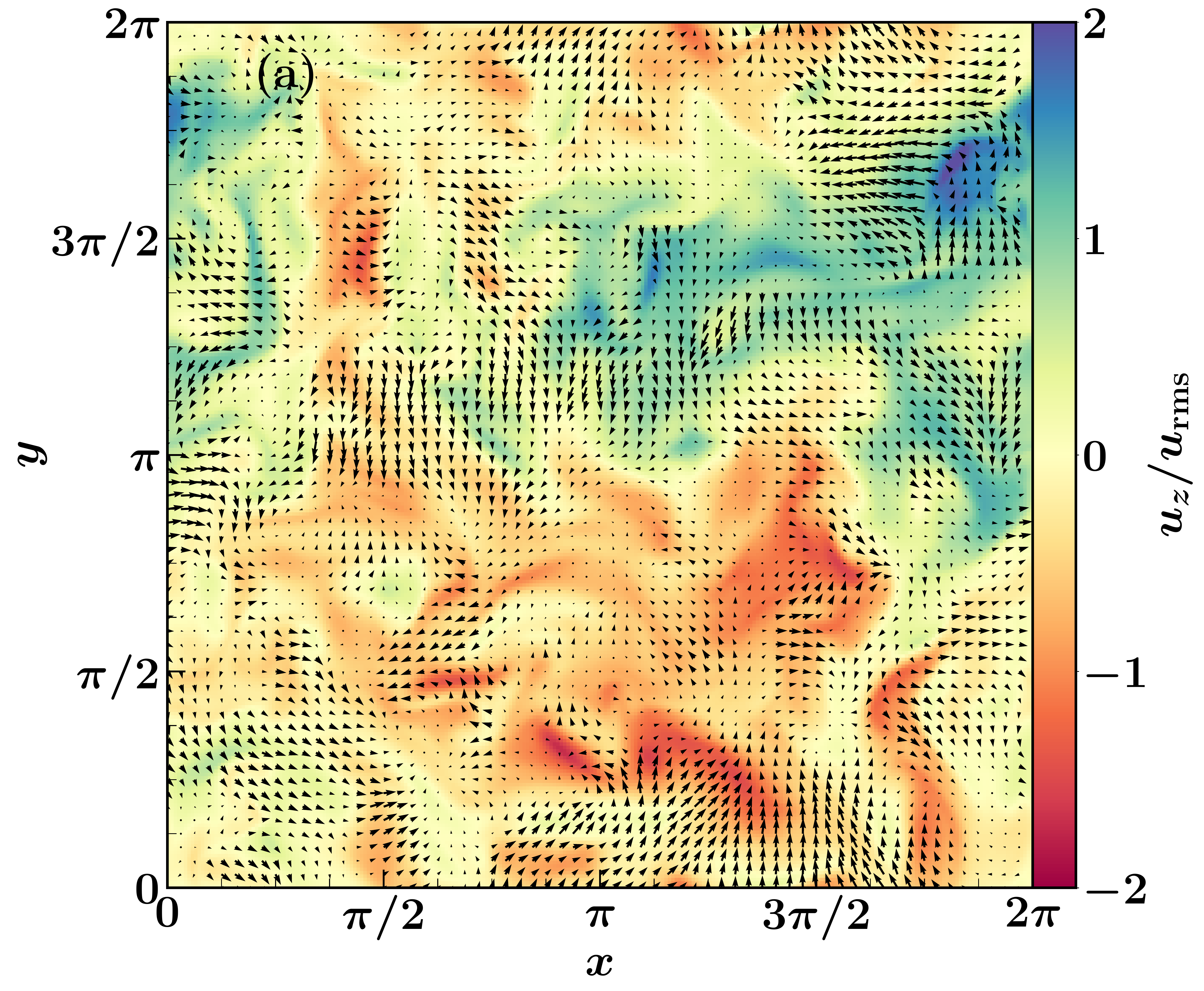}
\includegraphics[width=\columnwidth]{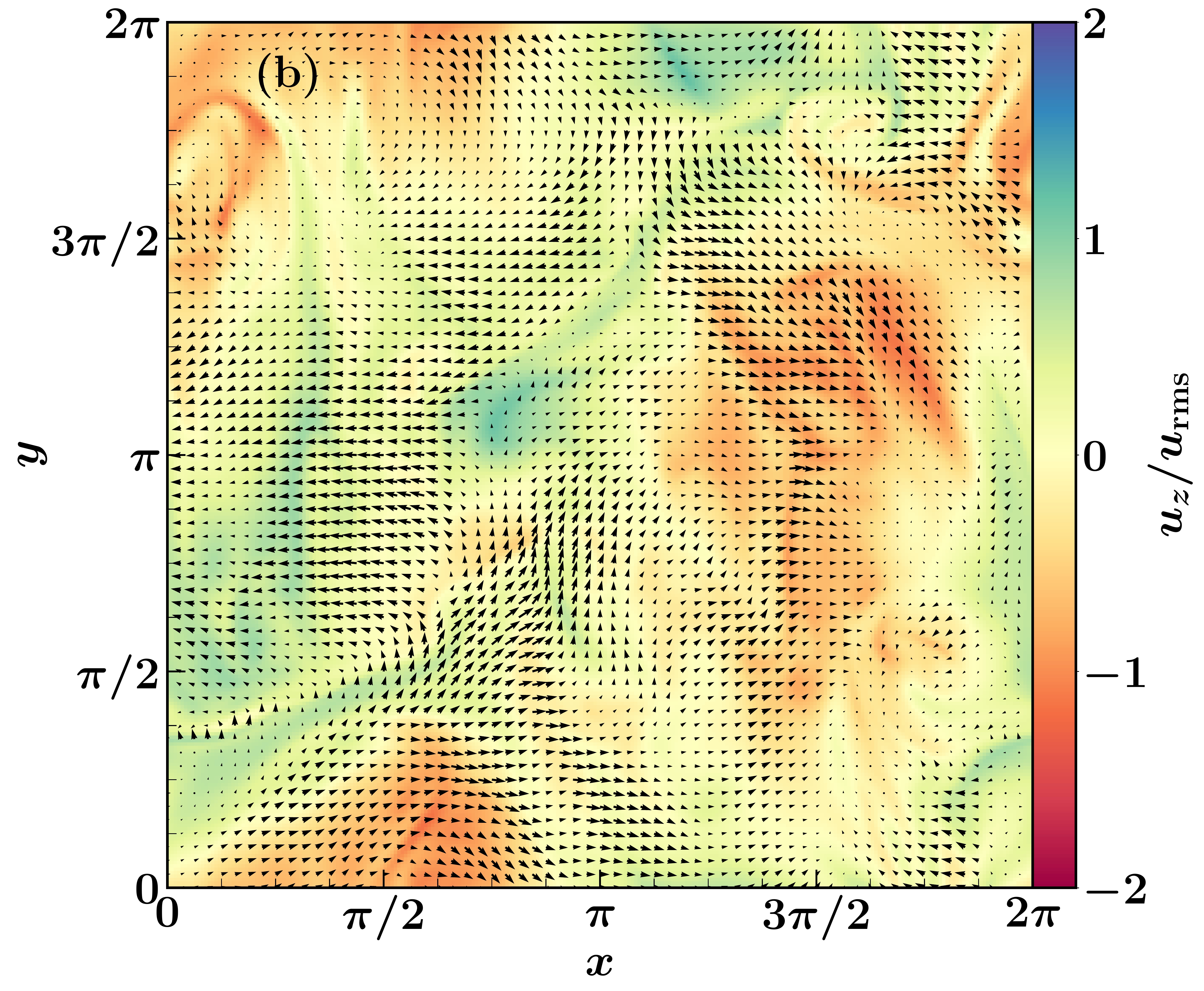}
	\caption{A 2D cut in the $xy$-plane with vectors $(u_x/\urms,u_y/\urms)$ and colour showing the magnitude of $u_z/\urms$ in
the kinematic (a) and saturated (b) stages with $\Rm=2244$. The velocity field in both the stages looks qualitatively the same. The structures
span approximately half of the domain.}
\label{2du} 
\end{figure*}
 \Fig{updf}
shows the PDF of a single component of the velocity field $u_x/\urms$ in the kinematic and saturated dynamo stages for $\Rm=1122$ and $\Rm=2244$.
The PDF is nearly Gaussian in both the kinematic and saturated stages. This is generally true for homogenous turbulence \citep{VM91}.
The velocity PDFs remain Gaussian even in the case of supersonic turbulence with a compressible forcing \citep[e.g., Fig. A1. in][]{Federrath2013}.
For all cases of \Tab{table_nfd}, the kurtosis of the velocity field is very close to ${\rm K}=3$, which is the value for a Gaussian distribution. 
The correlation length of the velocity field $\lu$, also given in \Tab{table_nfd}, is about half of the periodic domain size $L = 2 \pi$, as can also be seen from \Fig{2du}.
It decreases slightly as $\Re$ increases and is slightly larger in the saturated stage than in
the kinematic stage for all $\Rm$. The velocity field thus becomes more volume filling as the magnetic field saturates. 
This is directly attributable to the dynamical effects of the magnetic fields.

\begin{figure}
 {\includegraphics[width=\columnwidth]{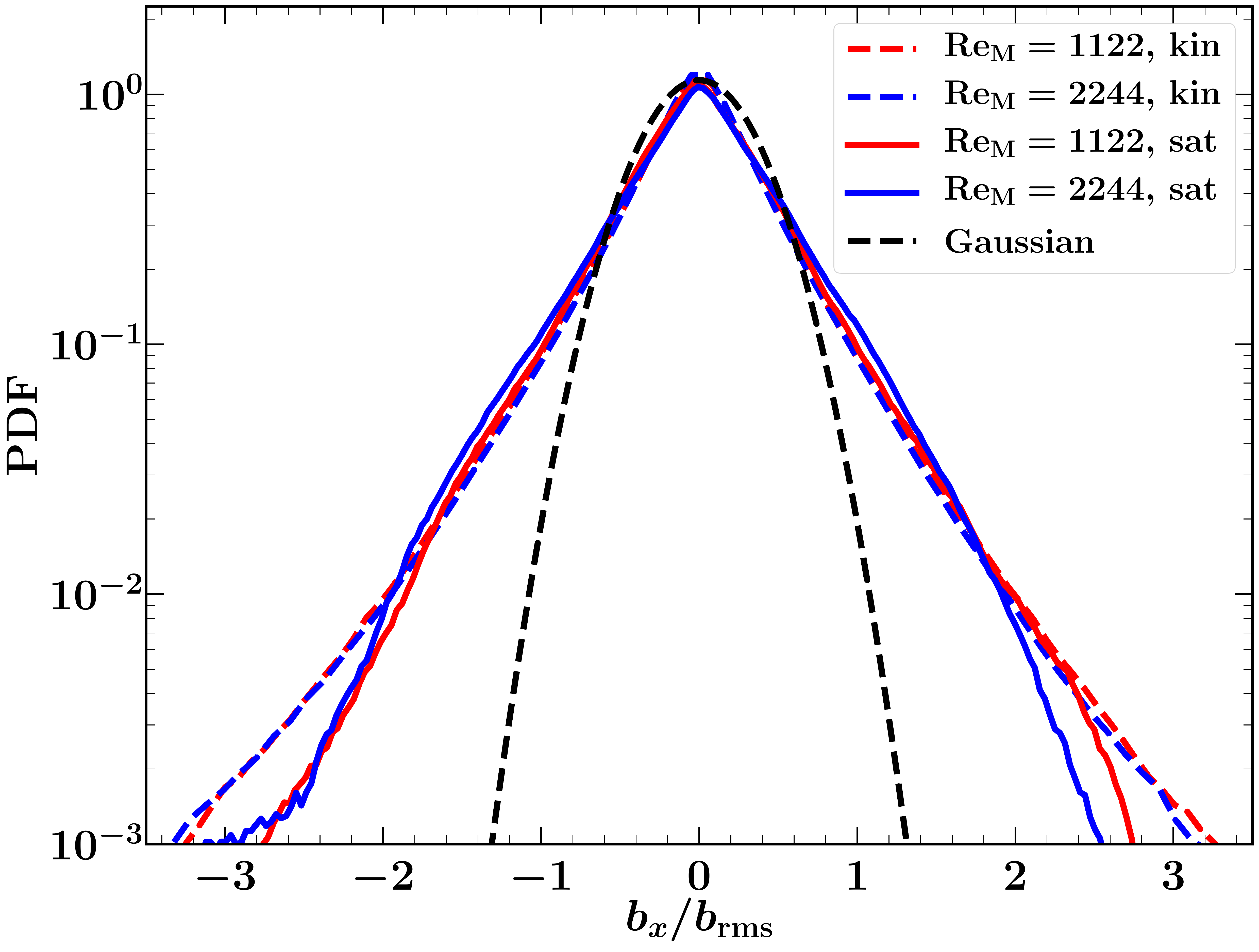}}
\caption{The PDF of the normalized magnetic field component $b_x/\brms$ for $\Rm=1122$ and $\Rm=2244$
in the kinematic (dashed) and saturated (solid) stages for the values of $\Rm$ given in the legend. The magnetic field for both $\Rm$ in both stages is far from a Gaussian (dashed, black). It has heavy tails, 
which is a sign of intermittency. Here only $b_x/\brms$ is shown but similar behaviour can be observed in all three magnetic field components.}
\label{bpdf} 
\end{figure}

\begin{figure}
 {\includegraphics[width=\columnwidth]{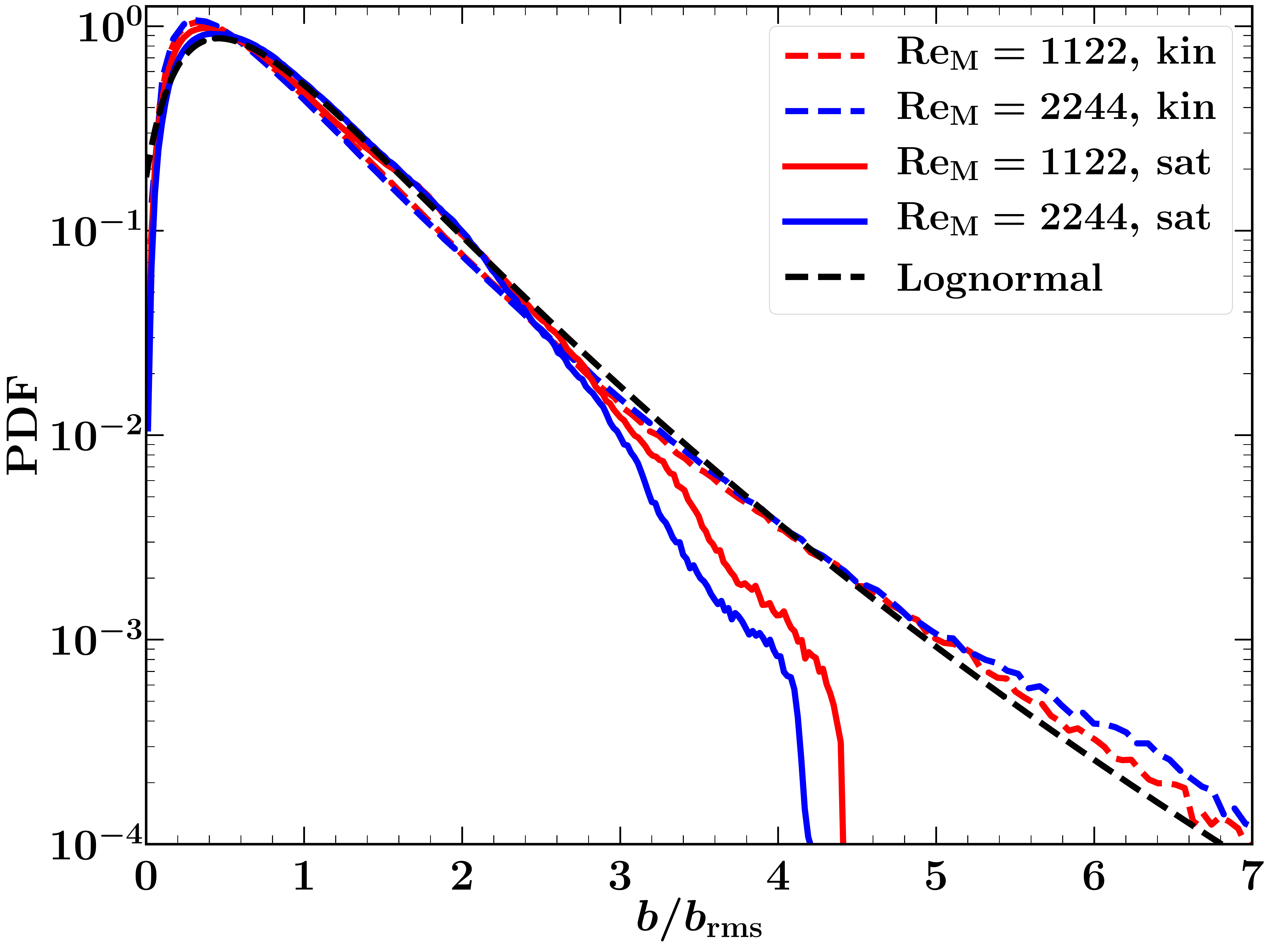}}
\caption{The PDF of the normalized magnetic field strength $b/\brms$ for $\Rm=1122$ and $\Rm=2244$ 
in the kinematic (dashed) and saturated (solid) stages for the values of $\Rm$ given in the legend. The PDF of the magnetic field in the kinematic state follows a lognormal distribution (dashed, black).
The magnetic field is more intermittent in the kinematic stage than in the saturated stage.}
\label{bmpdf} 
\end{figure}

\begin{figure*}
\includegraphics[width=\columnwidth]{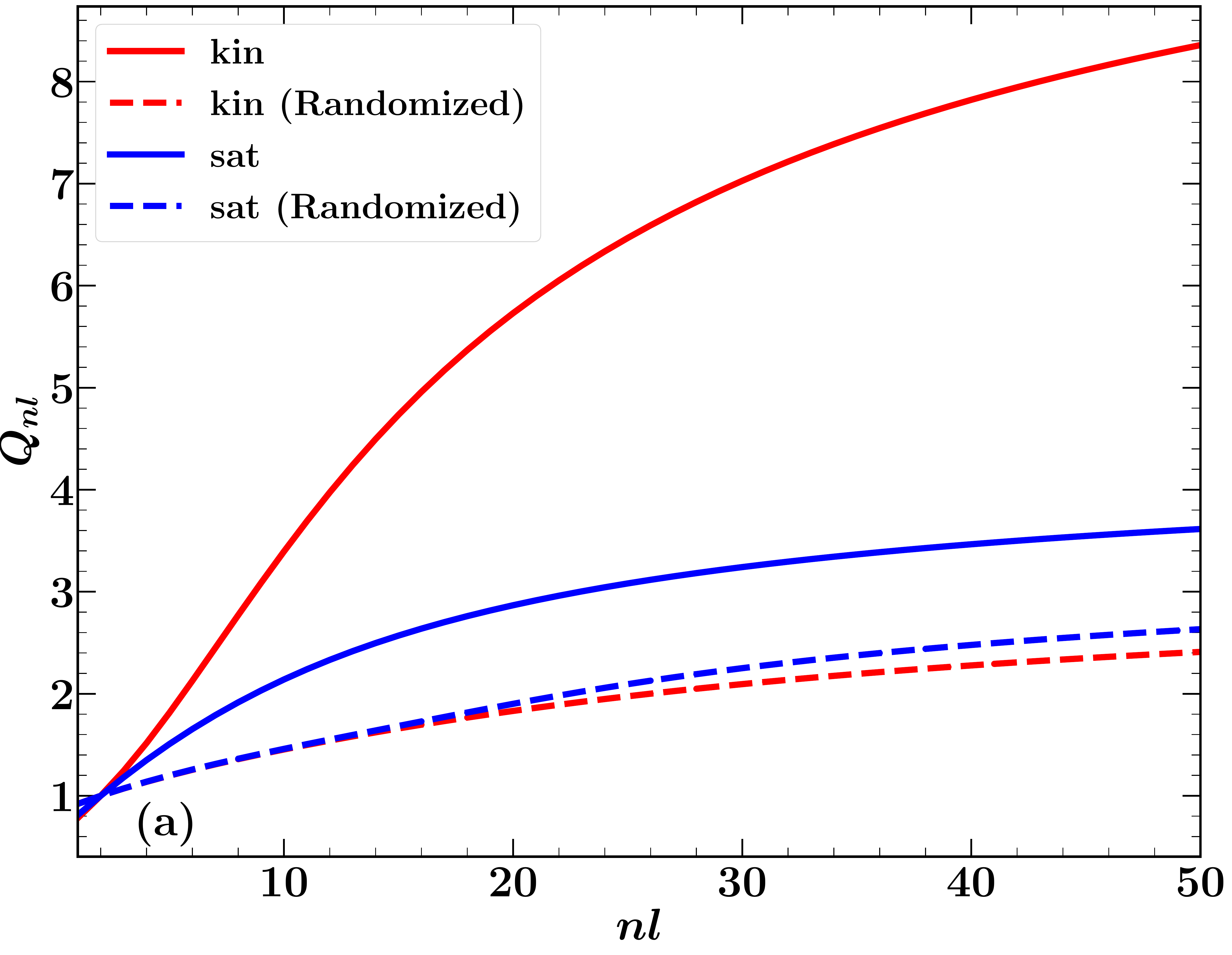}
\includegraphics[width=\columnwidth]{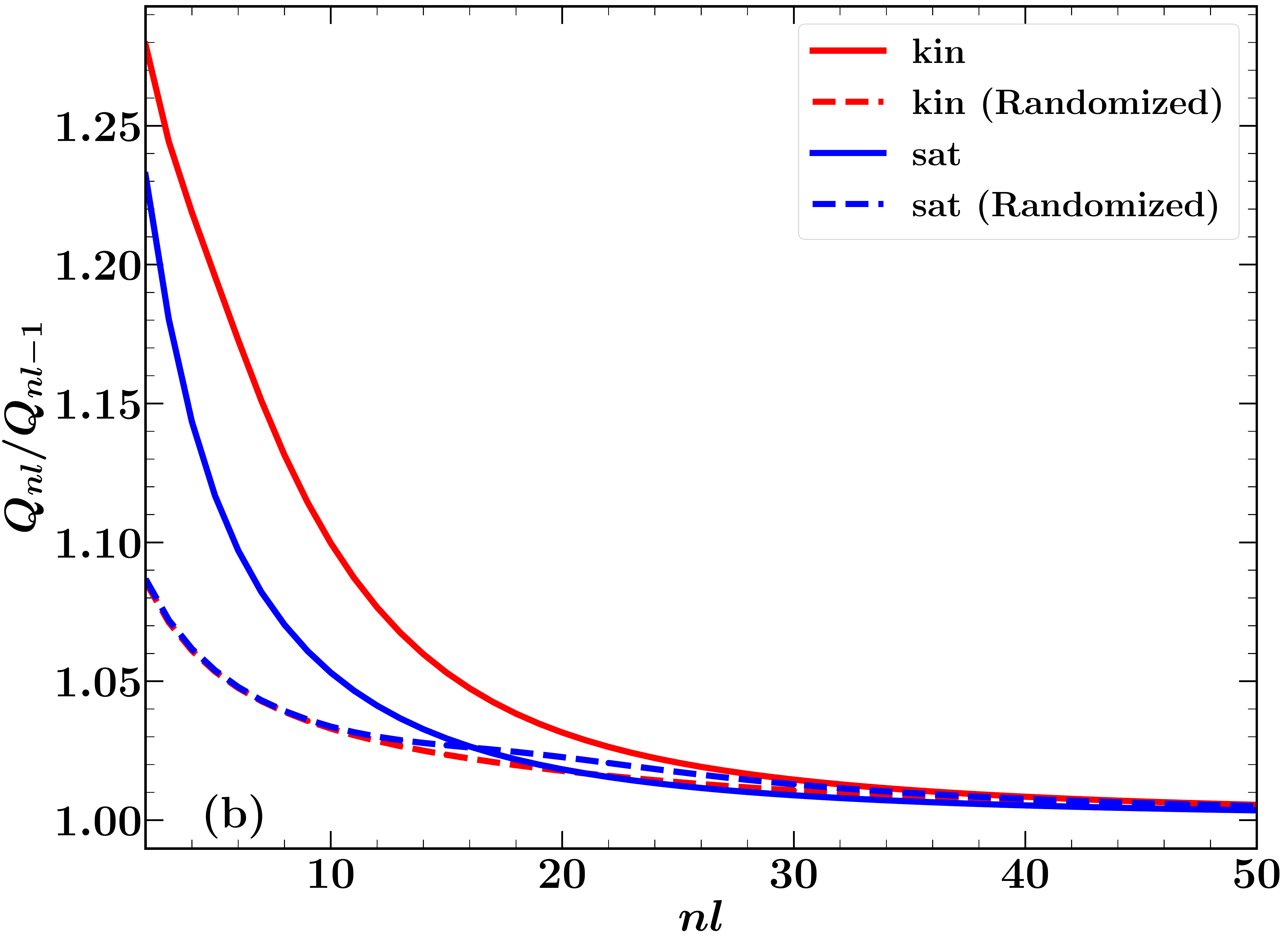}
	\caption{$Q_{nl} = \langle (b/b_{\rm rms})^{nl} \rangle ^{(1/nl)}$ (a) and $Q_{nl}/Q_{nl-1}$ (b) as function of $nl$ for the kinematic (red, solid) and saturated (blue, solid) stages for $\Rm=1122$. 
	The corresponding quantities for randomized fields which have almost Gaussian statistics (dashed) are also plotted. The dynamo generated magnetic field is always intermittent with the degree of intermittency being 
	higher in the kinematic stage.}
\label{qnl} 
\end{figure*}

\begin{figure*}
\includegraphics[width=\columnwidth]{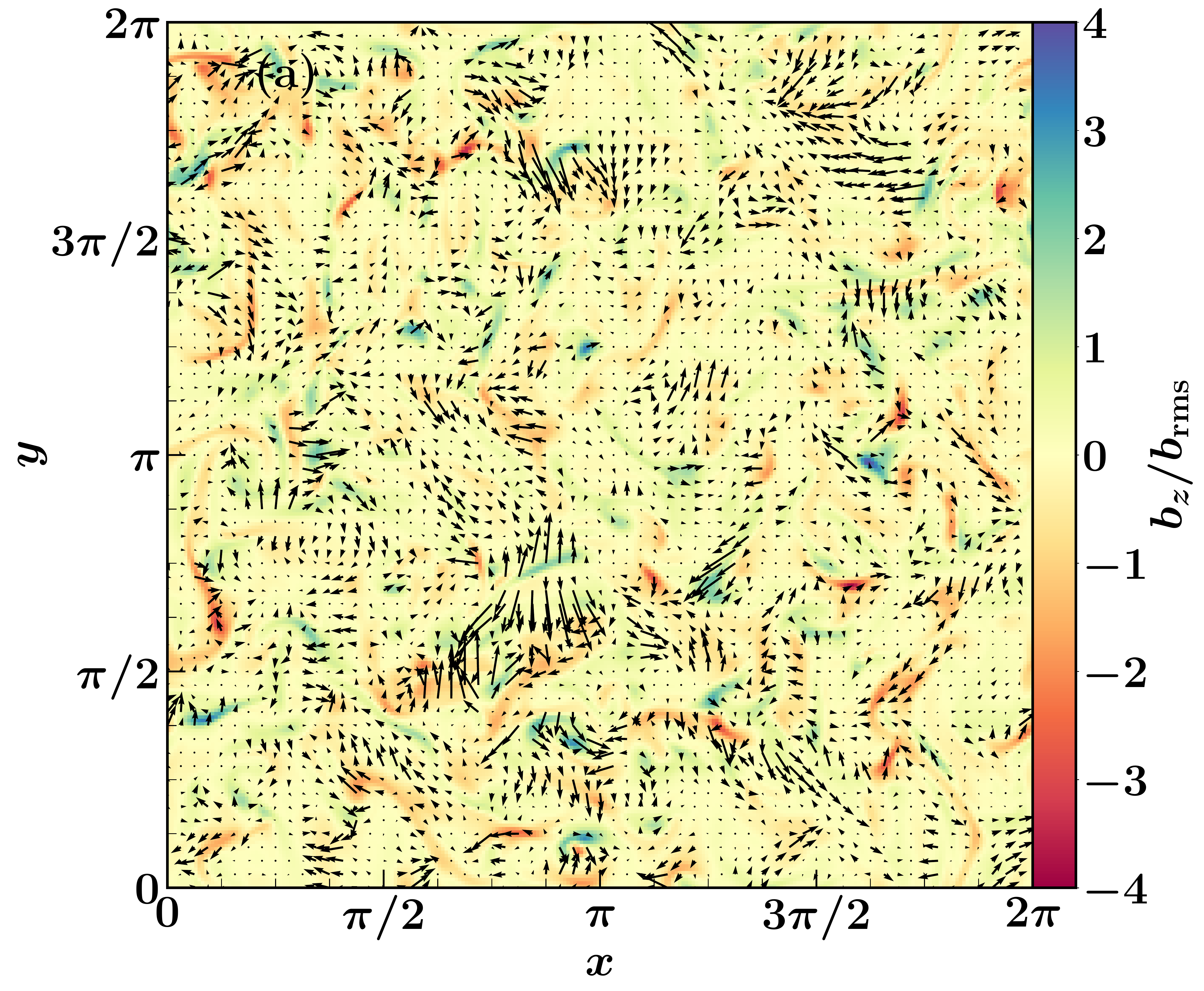}
\includegraphics[width=\columnwidth]{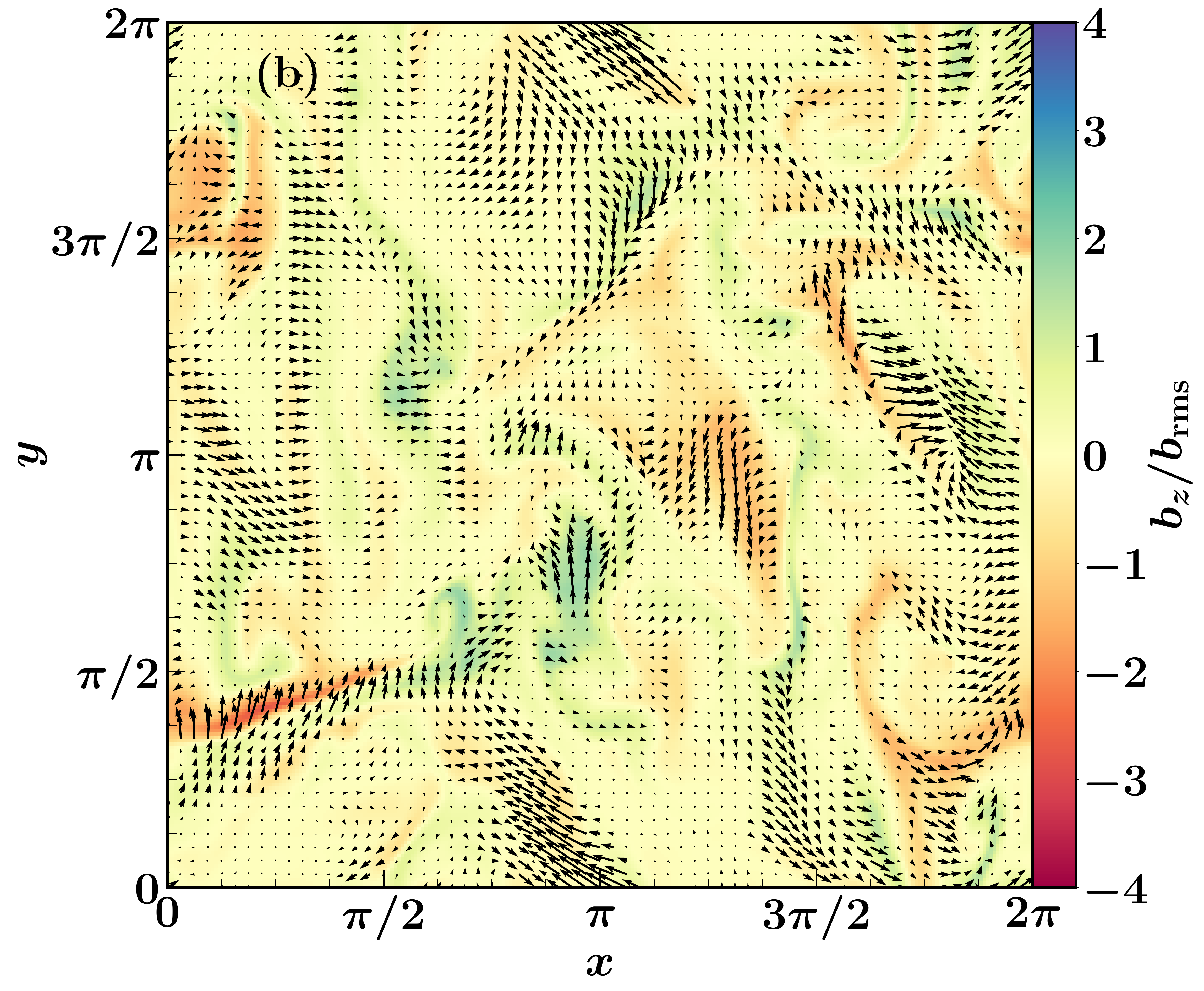}
	\caption{As \Fig{2du} but for the magnetic field. The magnetic field in the kinematic stage (a) is intermittent with random magnetic structures.
	In the saturated stage (b), the field remains intermittent but the structures are larger.}
\label{2db} 
\end{figure*}

\begin{figure*}
\includegraphics[width=\columnwidth]{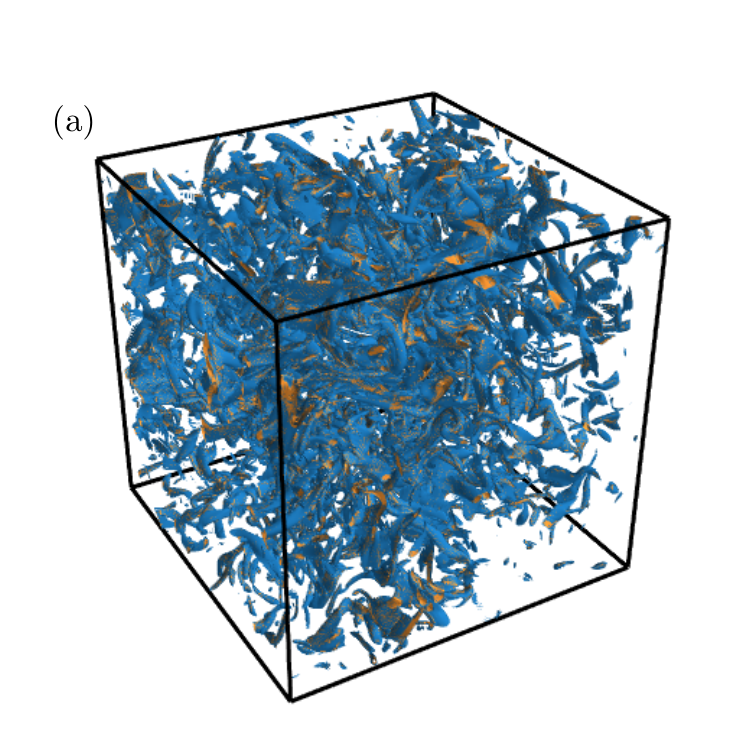}
\includegraphics[width=\columnwidth]{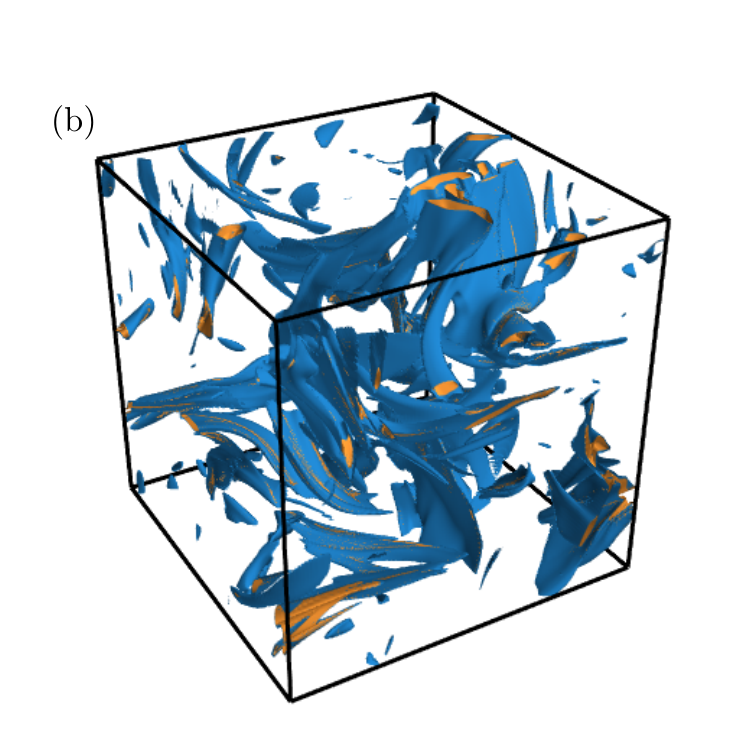}
\caption{Isosurfaces of $b^2/\brms^2=4$ (blue) and $b^2/\brms^2=5$ (yellow) for the magnetic fields in the kinematic (a) and saturated stages (b) for $\Rm=2244$.
The structures in the saturated stage are larger in size as compared to that in the kinematic stage.}
\label{3db} 
\end{figure*}

Even though the velocity field statistics are nearly Gaussian, the magnetic field in both the kinematic and saturated stages is spatially intermittent and strongly non-Gaussian. 
This can be seen from the PDFs of a normalized component of the magnetic field $b_x/\brms$ in \Fig{bpdf}. The distribution is far from a Gaussian one
and has long, heavy tails. The nonlinearity truncates the most extreme relative magnetic field strengths above $|b_x|/\brms\approx3$. 
The magnetic field intermittency is further demonstrated in
\Fig{bmpdf} which shows the PDF of $b/\brms$
for $\Rm=1122$ and $\Rm=2244$ in the kinematic and saturated stages. The PDF of the kinematic magnetic field strength follows
a lognormal distribution and it has heavier tails in comparison to that of the saturated magnetic field. 
Thus the magnetic field is intermittent in both the kinematic and the saturated stages, but the level of intermittency decreases as the field saturates. 
It should be noted that this conclusion is consistent with that of Schekochihin et al. \citep{SCTMM04} (\Fig{bmpdf} in this paper is similar to their Fig.~27), who studied a closely related system. This confirms that this finding is robust to small variations in the model setup and parameters.

Magnetic intermittency can also be quantified by measuring the quantity $Q_{nl} = \langle (b/b_{\rm rms})^{nl} \rangle ^{(1/nl)}$ and
its rate of change as $nl$ changes (for example, $Q_{nl}/Q_{nl-1}$). Higher $Q_{nl}$ and $Q_{nl}/Q_{nl-1}$  is a signature of a larger degree of intermittency.
\Fig{qnl} shows $Q_{nl}$ and  $Q_{nl}/Q_{nl-1}$ for the magnetic field in the kinematic and saturated stages for $\Rm=1122$ for $nl=1,2,3,\cdots,50$. 
$Q_{nl}$ and its rate of change are higher for the kinematic stage as compared to the saturated stage. This further demonstrates that the magnetic field in the saturated
stage is less intermittent than that in the kinematic stage. We further compare both terms with the corresponding Gaussian versions obtained
by randomizing phases in Fourier space \citep[keeping the exact same magnetic field spectrum but destroying intermittent structures, as done in][]{WSE09,SSSBW17,SSWBS18}. 
$Q_{nl}$ and $Q_{nl}/Q_{nl-1}$ are higher for the dynamo generated field in comparison to 
its randomized Gaussian versions in both the kinematic and saturated stages. Thus, the dynamo generated field is always spatially intermittent and
the degree of intermittency decreases as the field saturates due to nonlinearity.

The two-dimensional vector plots of the magnetic fields in \Fig{2db} also show larger
structures in the saturated stage. This can be further seen in \Fig{3db}, which shows the isosurfaces of magnetic fields
in the kinematic and saturated stages.
The kurtosis of the kinematic magnetic field for $\Rm=1122$ is $5.29$ but is $3.32$ in the saturated stage. This also suggests that the magnetic
field in the kinematic stage is more intermittent than the saturated stage.
The magnetic field correlation length $\lb$ is calculated using \Eq{lcu} by replacing $E_k$ with $M_k$, the magnetic field power spectrum. The magnetic field
correlation length in the kinematic ${\lb}_{\rm kin}$ and saturated ${\lb}_{\rm sat}$ stages is given in \Tab{table_nfd}. 
The magnetic field correlation length decreases as $\Rm$ increases, both for the kinematic and saturated stages (see \Sec{sec:morph} for further details). 
Thus, the magnetic field intermittency increases during both kinematic and saturated dynamo stages as $\Rm$ increases.
It is also clear that ${\lb}_{\rm sat} > {\lb}_{\rm kin}$ for all $\Rm$ which confirms again that the magnetic field in the kinematic stage is less volume filling. 
The increase in the correlation length due to magnetic field saturation is true regardless of the choice of $\Rm$ and agrees with previous numerical studies \citep{CR09,BS13}.

\section{Saturation of the fluctuation dynamo} 
\label{sec:sat}
Several mechanisms have
previously been
considered to explain the saturation of the fluctuation dynamo,
including a reduction in magnetic field line stretching due to the suppression 
of the Lagrangian chaos in the velocity field \citep{CHK96,Kim1999}, 
changes in the
mutual alignment of the velocity and magnetic field lines \citep{FB12},
the folded structure of magnetic fields and energy equipartition between magnetic and velocity fields for $\Pm \gg 1$  \citep{SCHMM02, SCTMM04},
enhancement in diffusion due to additional nonlinear velocity drift \citep{Sub99,Sub03} and
selective dissipation of the turbulent kinetic energy \citep{B65, MK02}.
From the induction equation \eqref{fdie}, there are two 
type of processes that could lead to the saturation: a decrease in the induction term $\left(\nabla \times (\vec{u} \times \vec{b})\right)$ or an increase in the
dissipation term $\left(\eta \nabla^2 \vec{b}\right)$. We explore each scenario here.

\subsection{Alignment of velocity field, magnetic field and electric current density}
\label{sec:sat1}
We first examine how the induction term is affected when the field becomes stronger. The rms magnitude of both the velocity and magnetic fields are statistically steady, as shown in \Fig{ts}. 
Thus, we consider the alignment of the magnetic field with the velocity field as a possible mechanism for the saturation. Such an alignment has been studied in the context of convectively driven 
fluctuation dynamos \citep{BJNRST96,FB12}, MHD turbulence in the presence of a strong guide field \citep{Mason06} 
and decaying isotropic MHD turbulence \citep{SMD08}. For the numerical simulations described in \Tab{table_nfd}, we calculate the
angle between the velocity $\vec{u}$ and magnetic field $\vec{b}$,
and between the current density $\vec{j}$ and $\vec{b}$,
\begin{align}
\aub = \frac{\vec{u} \cdot \vec{b}}{|\vec{u}||\vec{b}|},
\quad\mbox{and}\quad
\ajb = \frac{\vec{j} \cdot \vec{b}}{|\vec{j}||\vec{b}|},
\end{align}
respectively.  An increase in the level of alignment between $\vec{u}$ and $\vec{b}$ implies a decrease in the 
effectiveness of magnetic
induction. On the other hand,
an increase in the level of alignment between $\vec{j}$ and $\vec{b}$ leads to a decrease in the Lorentz force, i.e., the field becomes more force-free.

\begin{figure*}
\includegraphics[width=\columnwidth]{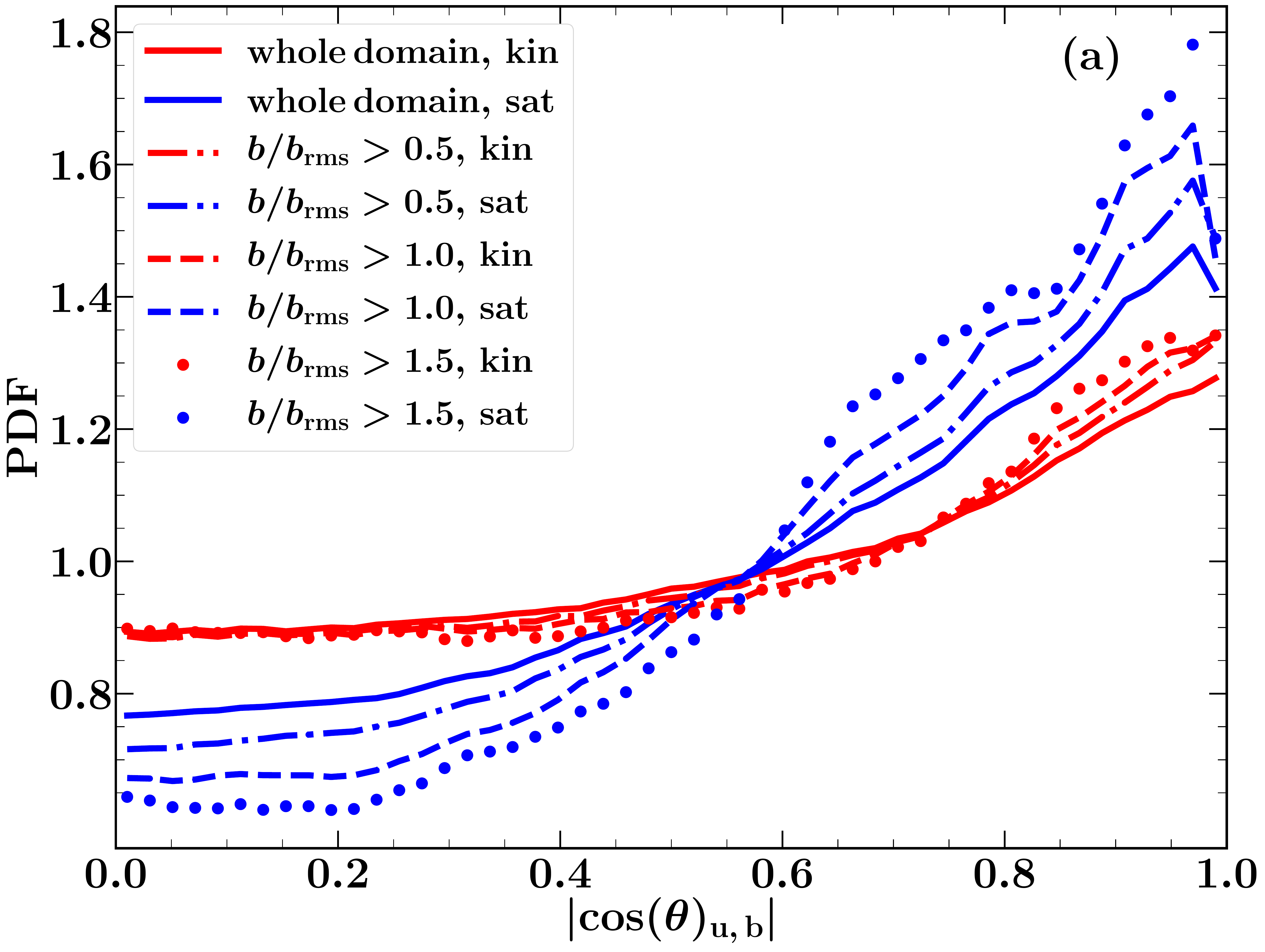}
\includegraphics[width=\columnwidth]{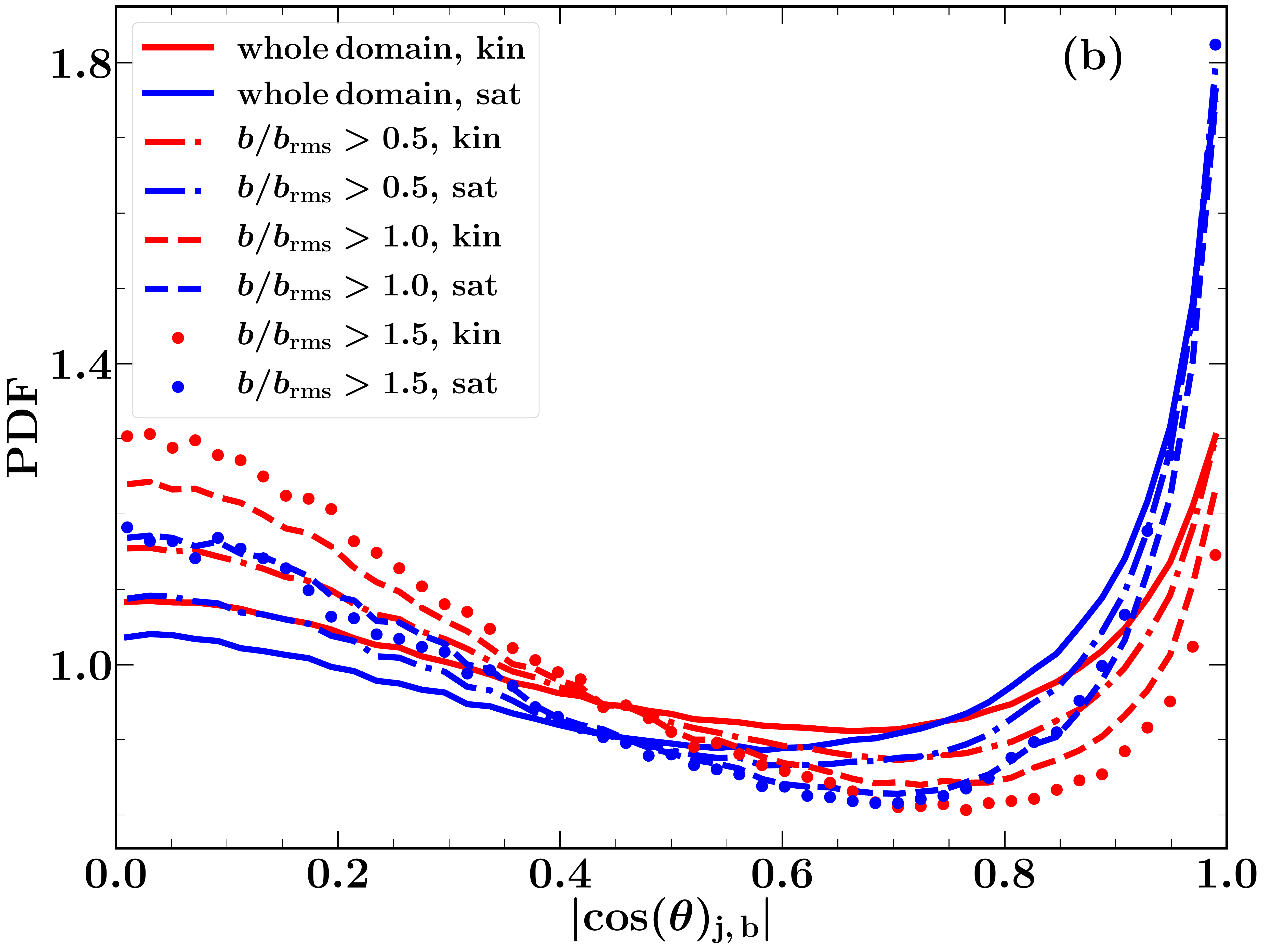}
\caption{The total and conditional probability distribution functions of the cosines of the angles between 
$\vec{u}$ and $\vec{b}$, $\aub$ (a) and between $\vec{j}$ and 
$\vec{b}$, $\ajb$ (b) for $\Rm=1122$ in the kinematic (red) and saturated (blue) states. The magnetic field 
in the saturated stage is more aligned with the velocity field (reducing the induction effects) 
as compared to the kinematic stage. The magnetic field 
also becomes better aligned with the electric current density, reducing the back reaction 
on the velocity field.}
\label{arun1} 
\end{figure*}

\begin{figure*}
\includegraphics[width=\columnwidth]{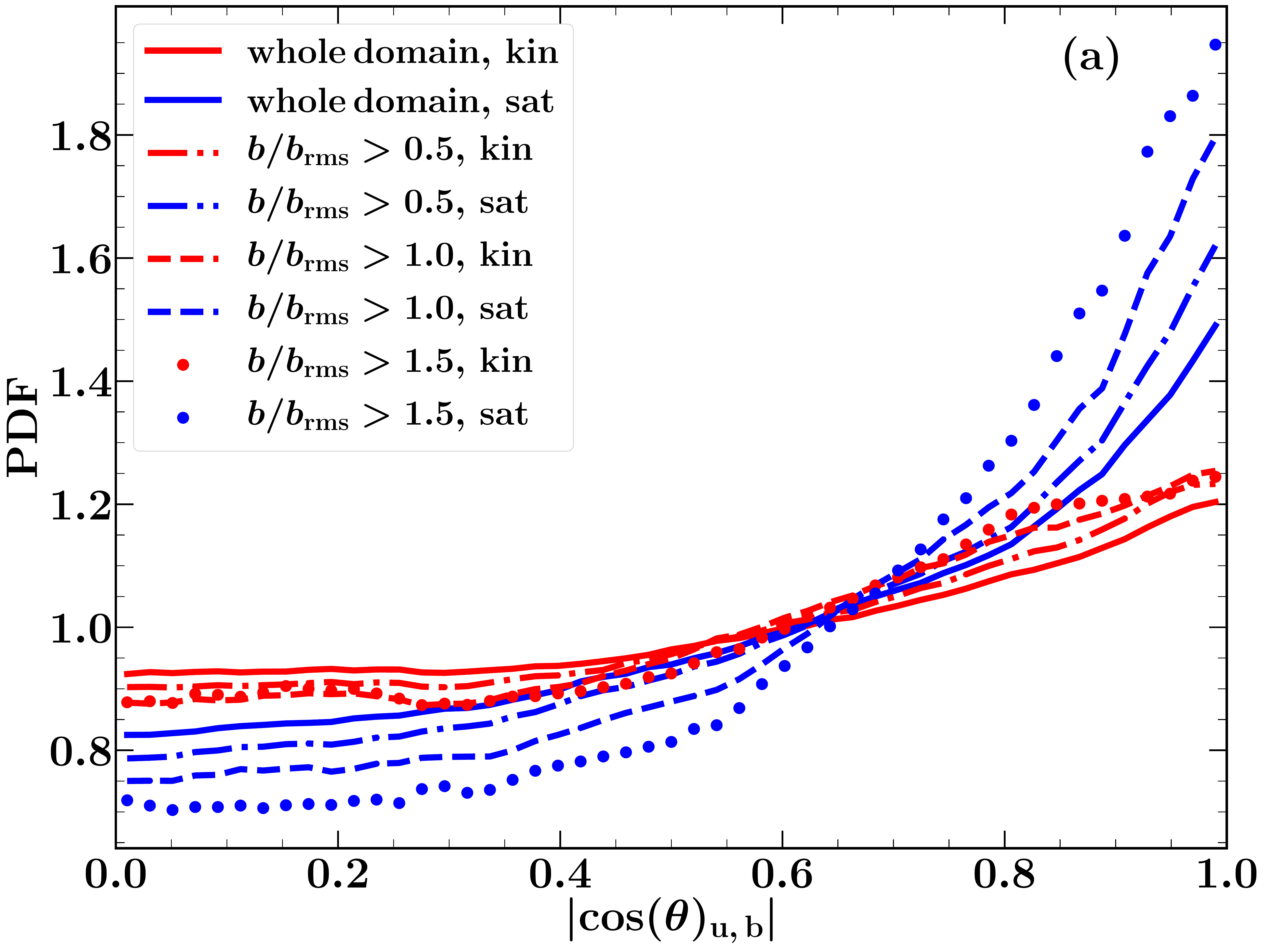}
\includegraphics[width=\columnwidth]{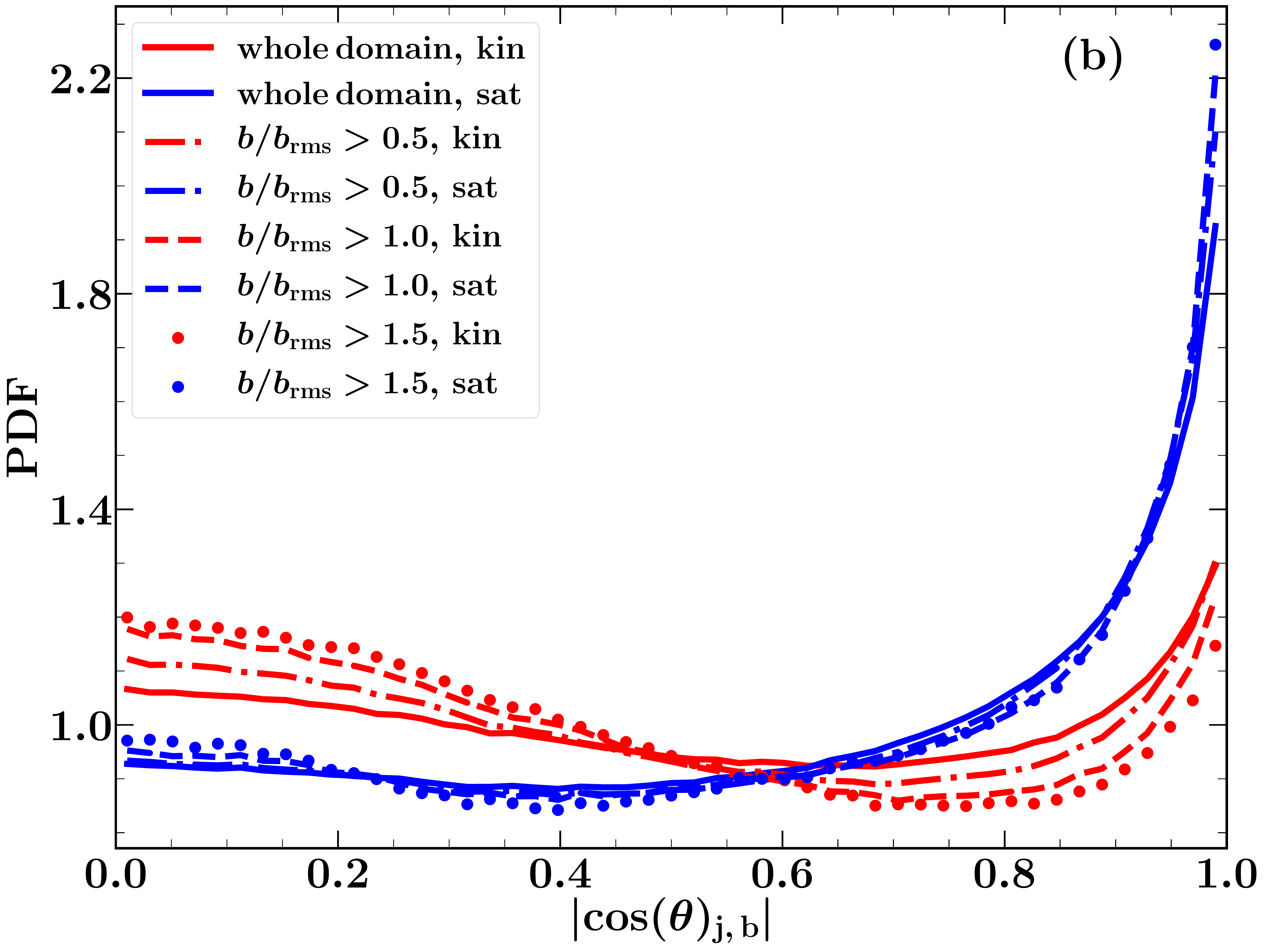}
\caption{As \Fig{arun1} but for $\Rm=1496$.}
\label{arun22} 
\end{figure*}

\Fig{arun1} and \Fig{arun22} show the probability density functions of the cosines in the kinematic and saturated stages for $\Rm=1122$ and $\Rm=1496$.
Since both angles are symmetric about $\vec{b}=\vec{0}$, we show PDFs of the absolute value of their cosines.
For both values of $\Rm$, the cosine of the angle between the velocity and magnetic field, $|\aub|$, tends to be larger in the saturated stage than in the kinematic stage. 
The better alignment between $\vec{u}$ and $\vec{b}$ decreases the induction term $\nabla \times (\vec{u} \times \vec{b})$ and thus reduces the amplification of the magnetic field. 
To put this another way, the enhanced alignment between $\vec{u}$ and $\vec{b}$ implies a decrease in the energy transfer from 
the flow to the magnetic field (which is a process that has been studied in some detail in the context of shell models of magnetohydrodynamic turbulence \citep{Verma04,Kumar13,Plunian13,Verma16}).
However, there is a significant fraction of the volume where the two fields are not aligned and so the amplification is not completely suppressed. This minimum level 
of amplification is required to balance the magnetic diffusion.  
The cosine of the angle between the current density and magnetic field $\ajb$ is also
statistically larger by magnitude in the saturated stage. Thus, the field becomes closer to a force--free form as it saturates. 
This also implies that the morphology of magnetic field changes on saturation, 
which motivates us to study the morphology of magnetic structures in \Sec{sec:morph}.
Overall, because of the enhanced local alignment between the velocity and magnetic field, the field amplification rate decreases. At the same time, due to the increase in the local alignment between the current density and
magnetic field, the field becomes more force--free.

Similar broad conclusions apply when we consider conditional PDFs that focus exclusively upon the regions of stronger field (higher $b/\brms$ in \Fig{arun1} and \Fig{arun22}). 
However, the level of alignment between the velocity and magnetic field is higher in the strong field regions in both the kinematic and saturated stages. This suggests
that the strong field regions require a larger reduction in amplification by alignment. The distribution of $\ajb$ in the kinematic stage shows some dependence
upon the field strength but in the saturated stage the difference is less pronounced.  
In the kinematic stage, alignment is weakest in the relatively strong field
regions, suggesting that in the strong field regions, not only because of its higher strength (as the Lorentz force is proportional to the strength of the field) 
but also because of the lower level of alignment, the field produces a stronger back reaction on the flow. 

\begin{figure*}
\includegraphics[width=\columnwidth]{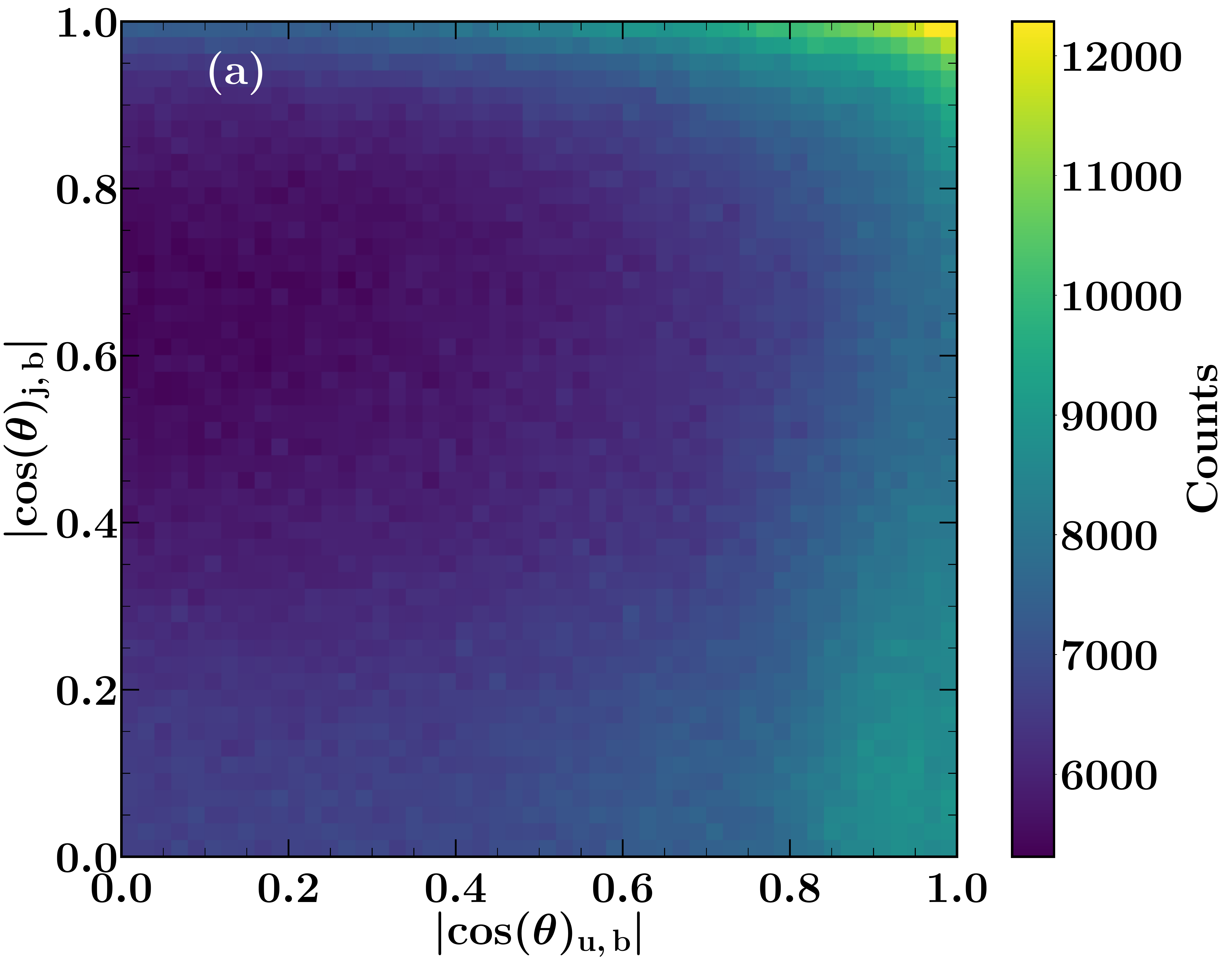}
\includegraphics[width=\columnwidth]{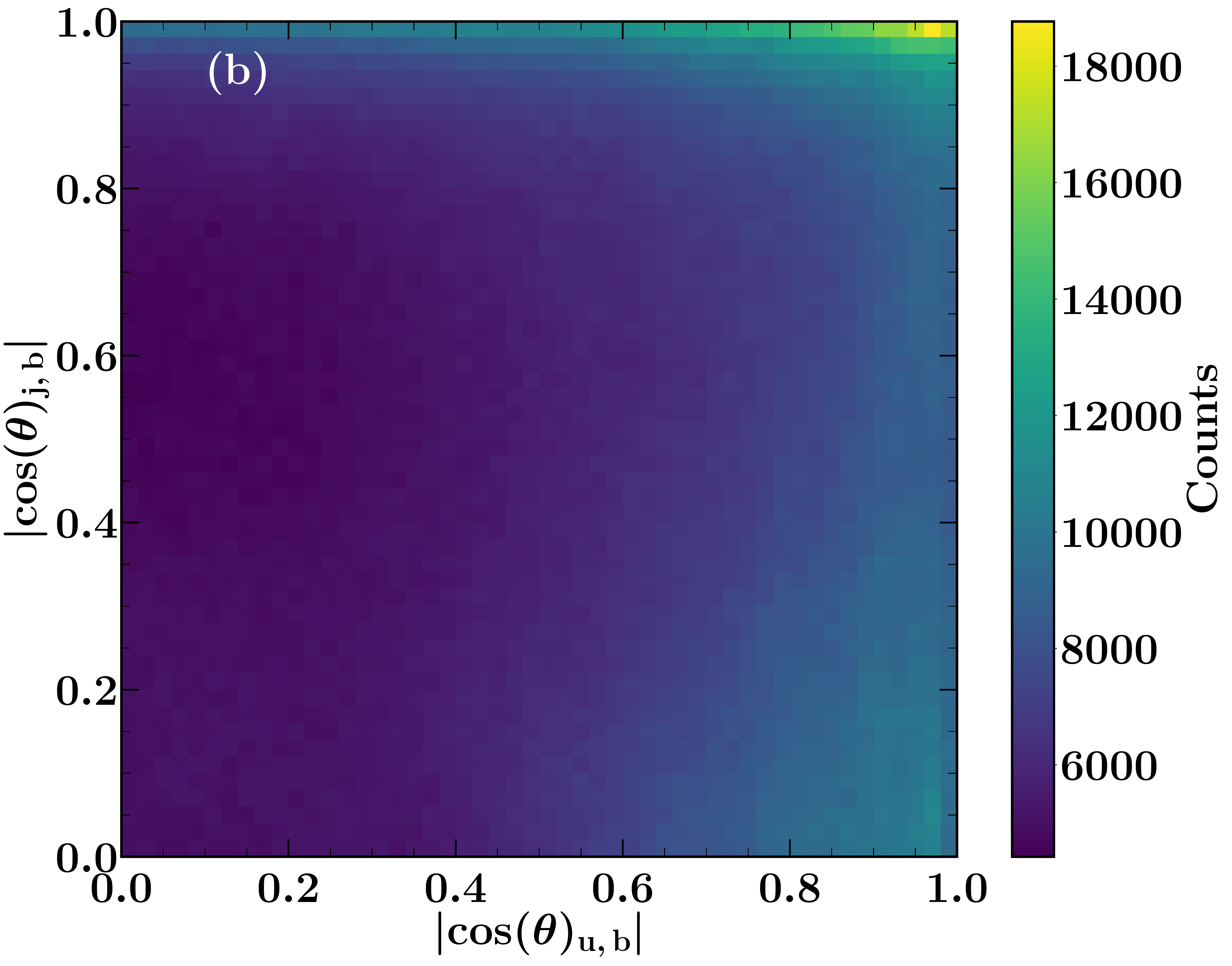}
\includegraphics[width=\columnwidth]{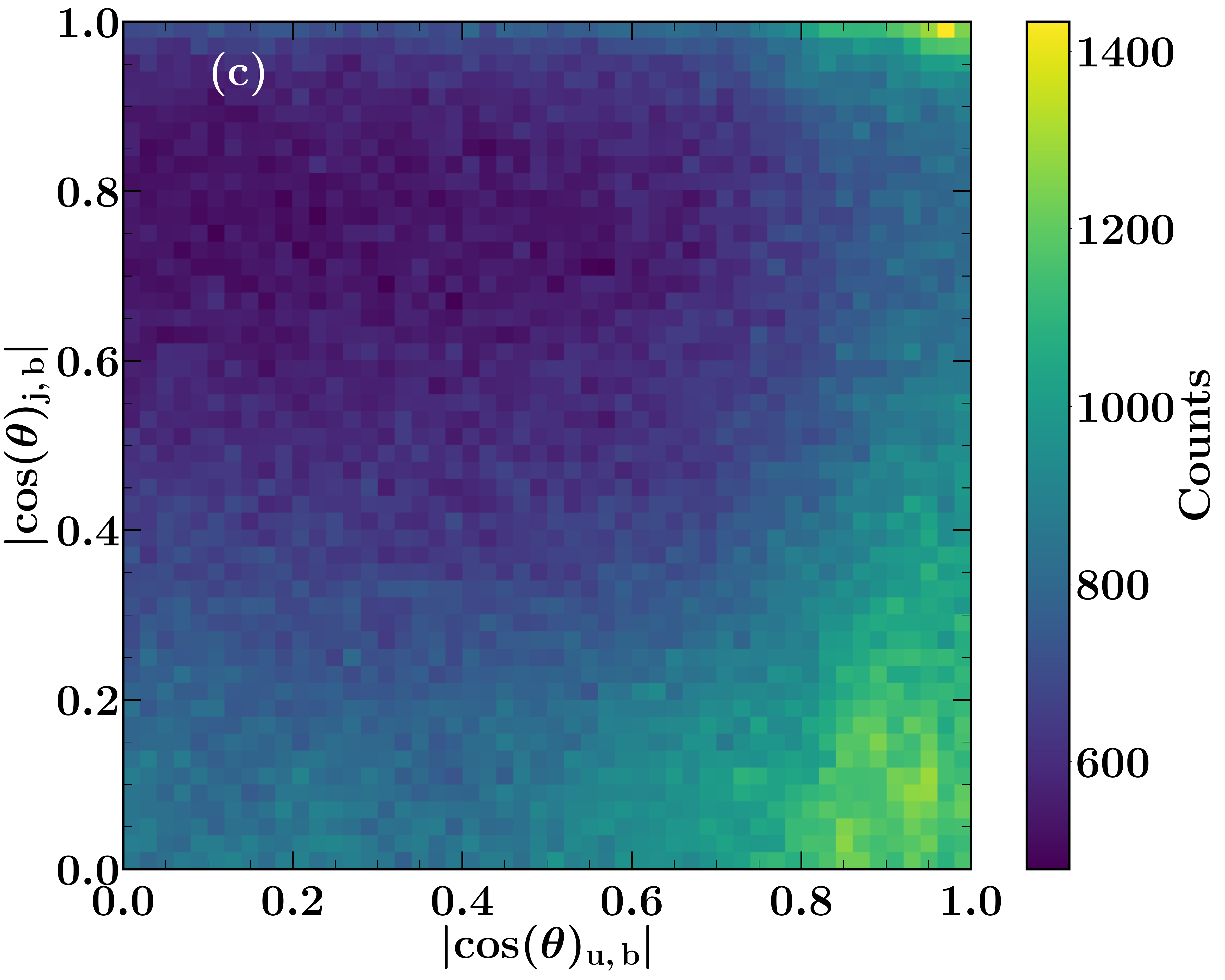}
\includegraphics[width=\columnwidth]{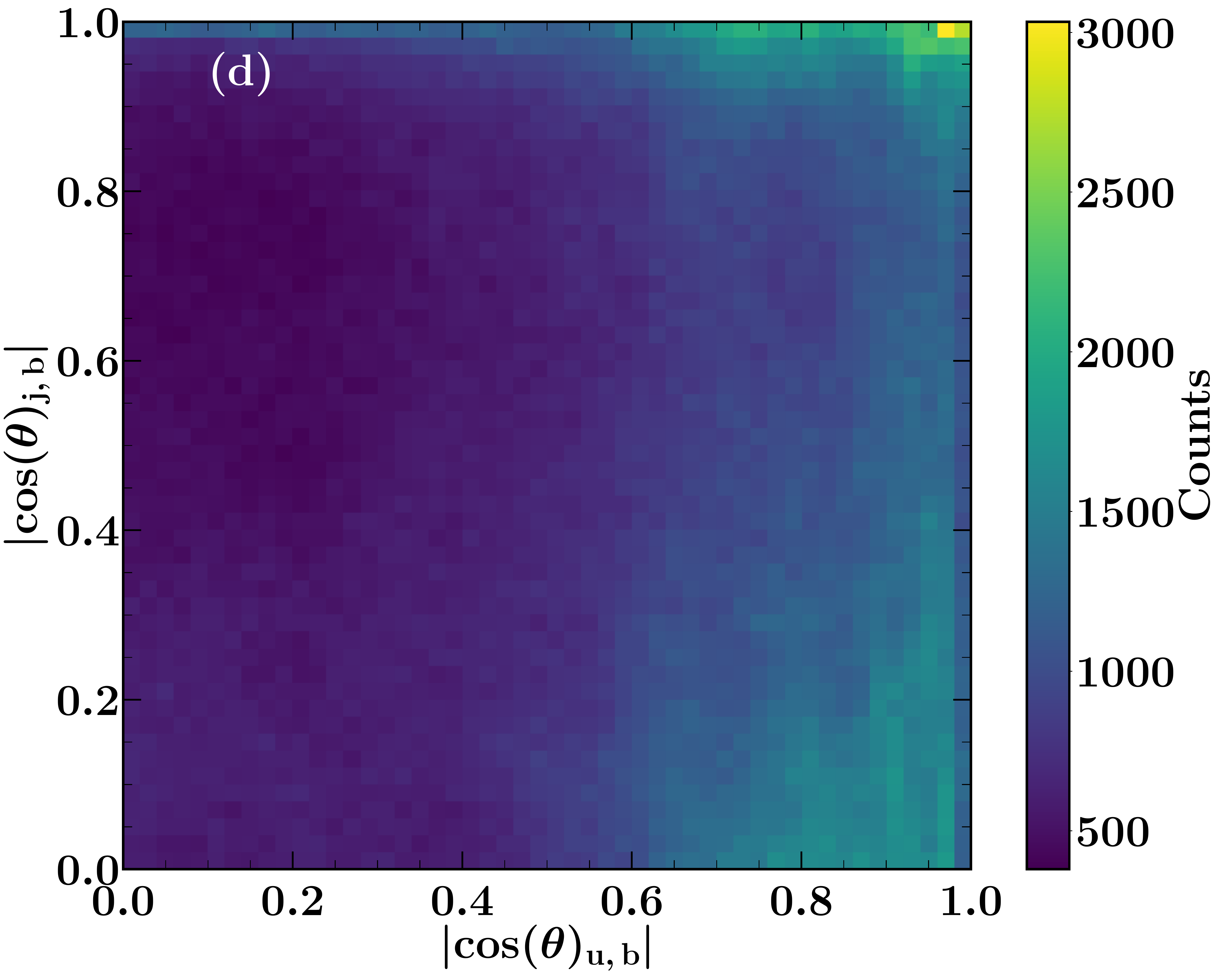}
\caption{The cross correlation of $\aub$ and $\ajb$ in the kinematic (a,c) and saturated 
(b,d) stages for  $\Rm=1122$. Panels (a) and (b) refer to the whole domain
 and the difference between them is not significant. Panels (c) and (d) refers to 
 only the strong field regions ($b/\brms \ge 1.5$). The yellow patch 
close to low $\ajb$ and high $\aub$ in the kinematic stage vanishes for the saturated stage. 
The peak in the count is always at high $\aub$ and high $\ajb$, which 
implies significant alignment between magnetic field, velocity field and current density. }
\label{ca} 
\end{figure*}

Another important question is whether the alignment between the velocity and magnetic fields and the magnetic field and current density occur in the same
spatial region. To answer this, we show the cross-correlation between the two angles in \Fig{ca} which
suggests that the velocity, magnetic field and current density are always nearly aligned to each other at same spatial positions. It is difficult to see any further difference between
the kinematic and saturated stages in \Fig{ca}a and \Fig{ca}b. \Fig{ca}c and \Fig{ca}d show the same correlation but only for strong field regions, $b/\brms > 1.5$. 
In \Fig{ca}c, the kinematic stage shows higher correlation in regions with high $\aub$ and low $\ajb$, which is absent in the saturated stage. 
The larger misalignment of $\vec{j}$ and $\vec{b}$, especially
in the strong field regions, enhances the work done on the magnetic field by the flow. This promotes growth of the magnetic field.
Once the field saturates, the larger correlation at high $\aub$ and low $\ajb$ disappears in \Fig{ca}d.
This implies a statistical decrease in the back-reaction of the magnetic field on the flow as the field saturates. 

To summarize, the alignment between the velocity and magnetic field vectors and the magnetic field and current density vectors is statistically enhanced as the dynamo saturates.
The alignment does not completely inhibit the amplification, so there is always some field generated to balance the resistive decay. This in turn also
implies that the back reaction of the Lorentz force always remains significant.

\subsection{Magnetic field stretching}
\label{sec:sat2}
\begin{figure*}
\includegraphics[width=\columnwidth]{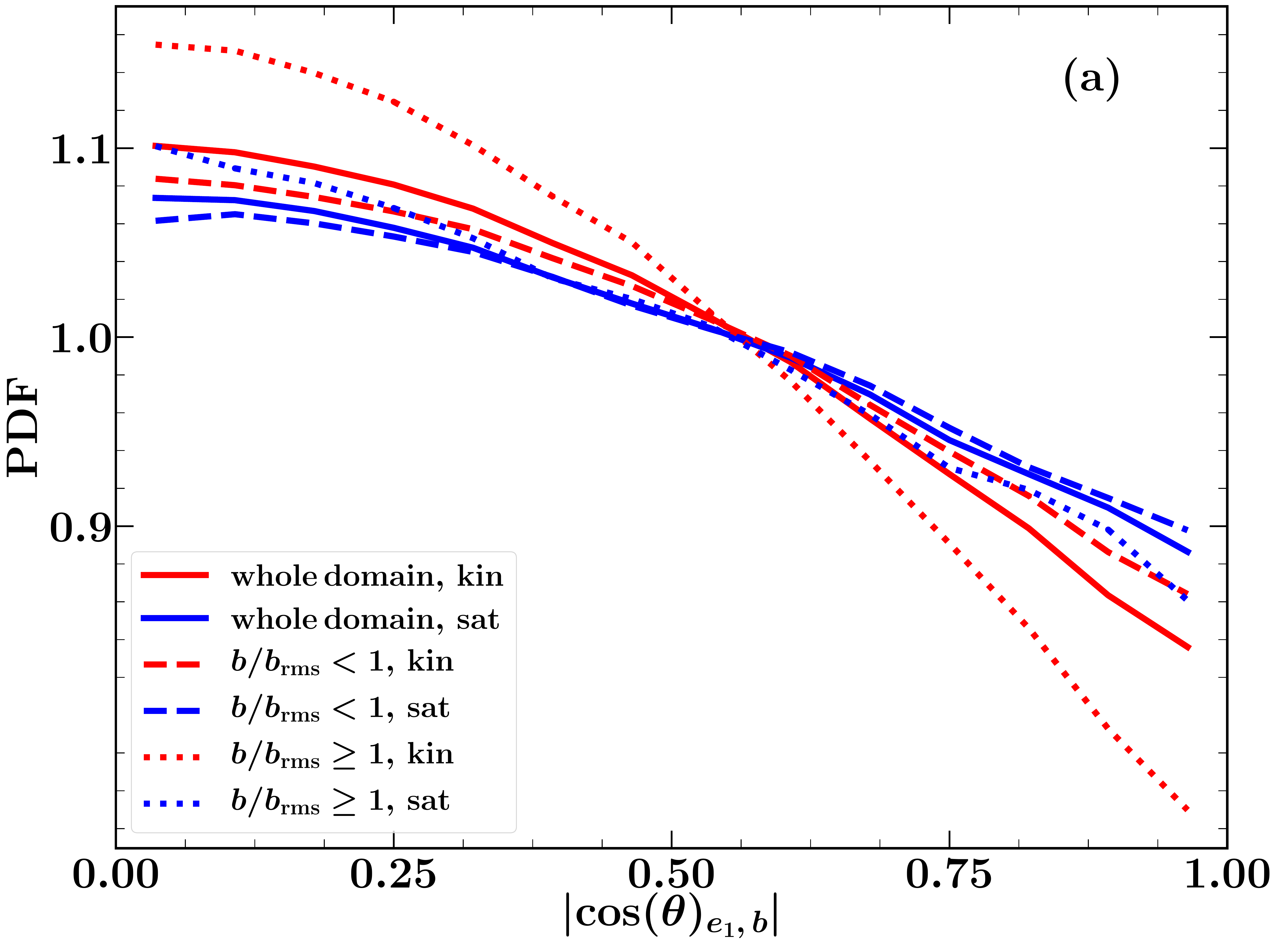}
\includegraphics[width=\columnwidth]{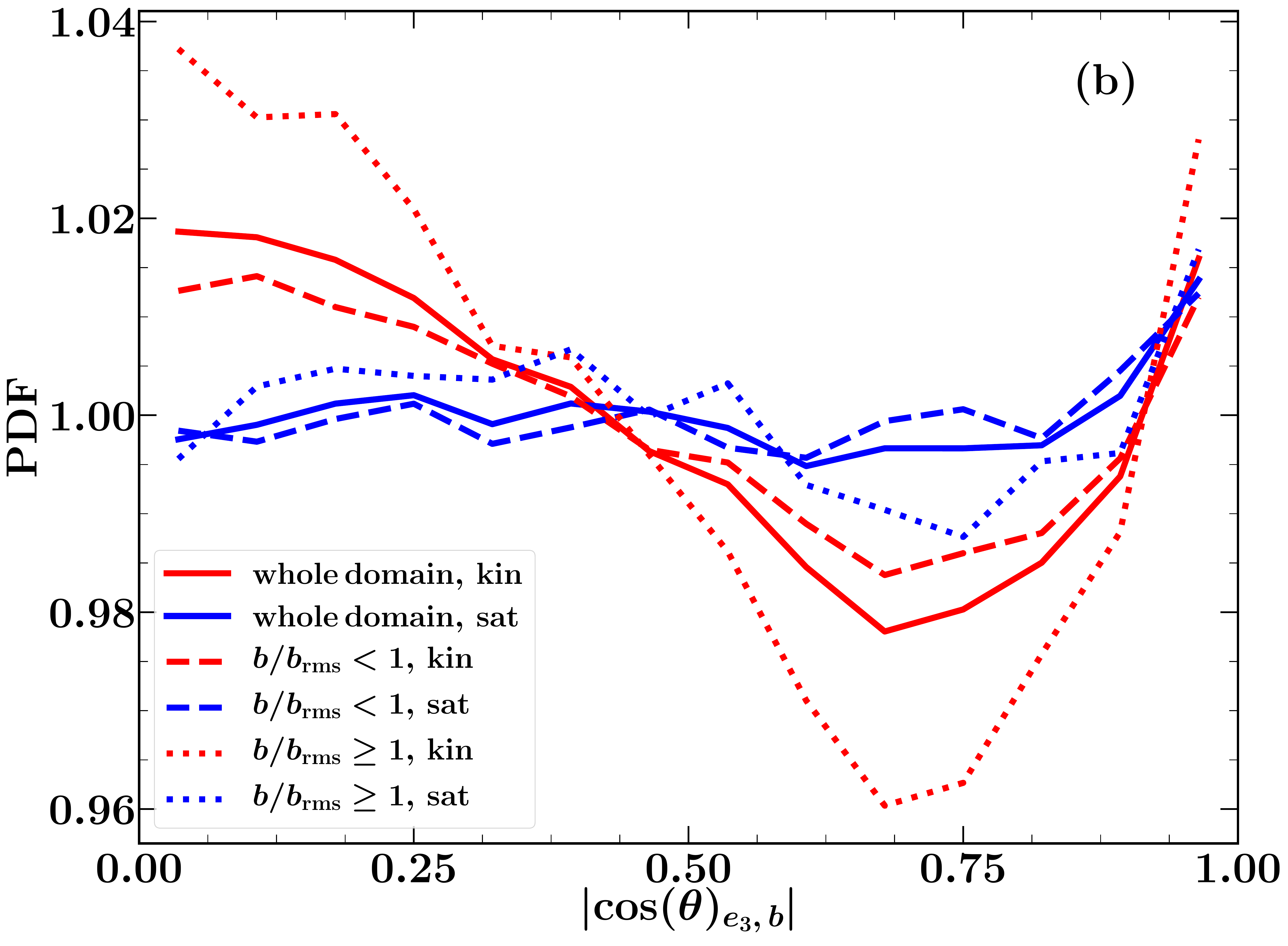}
\caption{The total and conditional PDFs of the cosine of the angle between the direction of local field line compression and the magnetic field $\aeob$ (a) 
and between the direction of local field line stretching and the magnetic field $\aetb$ (b) for $\Rm=1796$ in the kinematic (red) and saturated (blue) stages.}
\label{aerun25} 
\end{figure*}

\begin{figure*}
\includegraphics[width=\columnwidth]{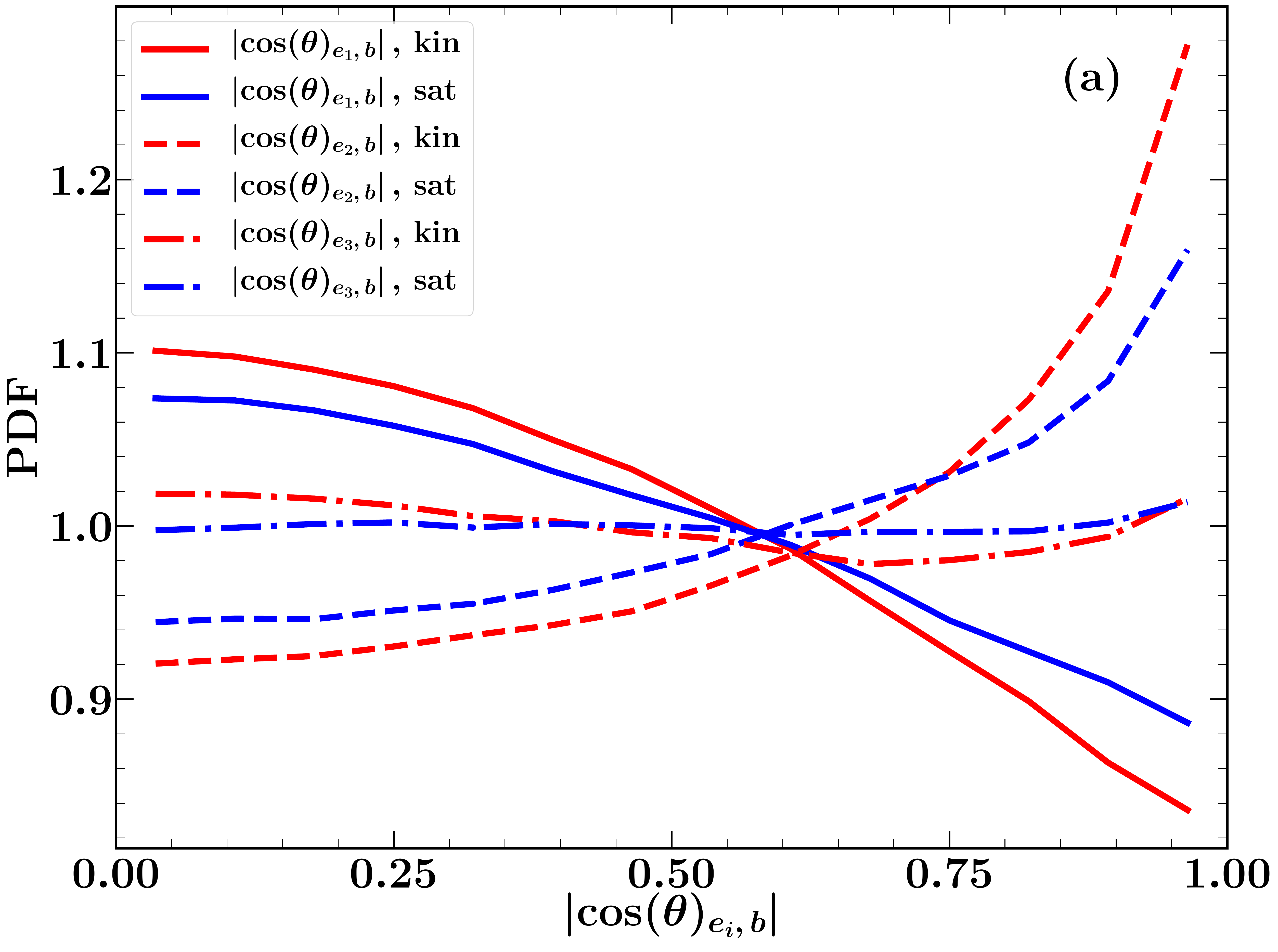}
\includegraphics[width=\columnwidth]{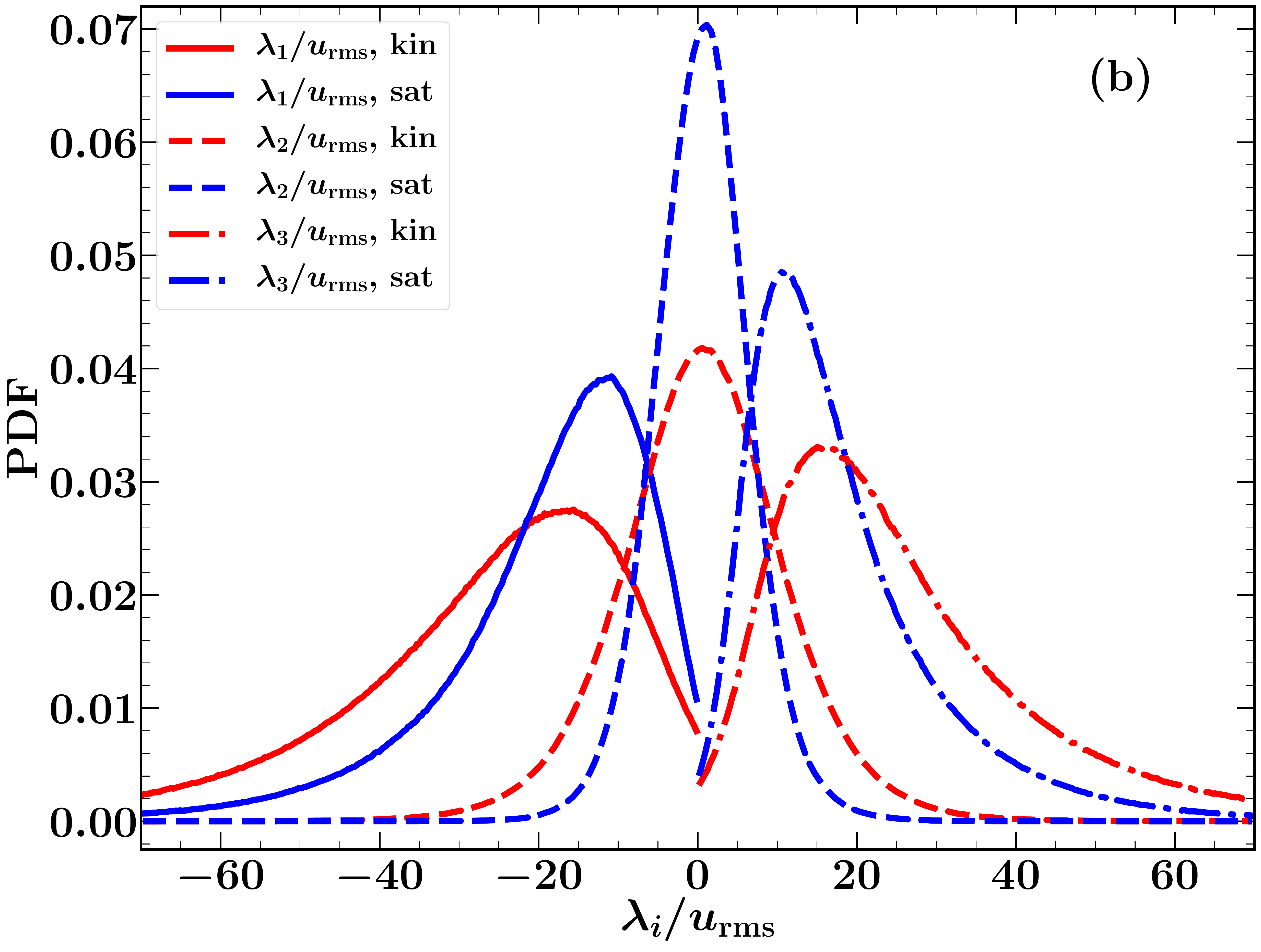}
\caption{The PDFs of the cosine of the angle between three three eigenvectors ($\vec{e}_1, \vec{e}_2,$ and $\vec{e}_3$) 
with the local magnetic field direction (a) and three three eigenvalues ($\lambda_1, \lambda_2,$ and $\lambda_3$) normalized by $\urms$ (b) for $\Rm=1796$ in the kinematic (red) and saturated (blue) stages.}
\label{lrun25} 
\end{figure*}

To explore another mechanism by which magnetic field amplification can be suppressed, we consider the stretching of the magnetic field lines by  the
turbulent velocity. For this, we consider the alignment of the magnetic field with the eigenvectors of the rate of strain tensor.  
Neglecting the rather weak divergence of the flow, the symmetric $3 \times 3$ matrix $S_{ij} = \frac{1}{2} \left(u_{i,j} + u_{j,i}\right)$ is calculated at each point in the domain
using sixth-order finite differences, and its eigenvalues and eigenvectors are calculated. The eigenvalues
are arranged in an increasing order, $\lambda_1 < \lambda_2 < \lambda_3$. The corresponding eigenvectors are $\vec{e}_1,\vec{e}_2,\vec{e}_3$.
The sum of the eigenvalues is close to zero since the flow is nearly incompressible. $\lambda_1$ is always negative and the vector $\vec{e}_1$
corresponds to the direction of local compression of magnetic field,  $\lambda_3$ is always positive and the vector $\vec{e}_3$
corresponds to the direction of local stretching, whereas  $\lambda_2$ can be obtained from $\lambda_1 + \lambda_2 + \lambda_3 \approx 0$.
The direction $\vec{e}_2$ (sometimes referred to as the `null' direction \citep{SCTMM04,SO19}) can correspond to either local stretching or compression depending on the sign of $\lambda_2$.
We then quantify the alignment with the magnetic field $\vec{b}$ of the vectors $\vec{e}_1$ and $\vec{e}_3$ by considering
\begin{align}
\aeob = \frac{\vec{e}_1 \cdot \vec{b}}{|\vec{e}_1||\vec{b}|}
\quad\mbox{and}\quad
\aetb = \frac{\vec{e}_3 \cdot \vec{b}}{|\vec{e}_3||\vec{b}|}.
\end{align}
\Fig{aerun25} shows the PDF of the cosines in the kinematic and saturated stages for $\Rm=1796$. In most of the volume, the direction 
of the magnetic field is perpendicular to the direction of the local compression (\Fig{aerun25}a), which leads to the amplification of magnetic field, and this trend is slightly stronger in the kinematic stage.
The PDF of the angle between the direction of local stretching and the magnetic field $\aetb$ has 
maxima at $\aetb=0$ and $\aetb=1$ in the kinematic stage. In the saturated stage, however, all angles are nearly equiprobable. This 
change in behaviour is more pronounced in the strong field regions, $b/\brms \ge 1$.
In \Fig{lrun25}a, we also show the PDF of $\aetwob$, $\aeob$ and $\aetb$.  
The forms of the PDF for $\aetwob$ are different from that of $\aeob$ and $\aetb$ in the kinematic and saturated stages.
The magnetic field is less aligned to the direction $\vec{e}_2$ in the kinematic stage as compared to the saturated stage and
its effect, locally on the magnetic field, is decided by the sign of the eigenvalue $\lambda_2$ (dashed lines in \Fig{lrun25}b). 
\Fig{lrun25}b shows the PDF of all three eigenvalues in the kinematic and saturated stages. All three eigenvalues are statistically lower in magnitude in the saturated stage 
as compared to the kinematic stage.
However, as can be seen in \Fig{aerun25} and \Fig{lrun25}a, the difference between the PDFs in the kinematic and saturated stages, whilst statistically significant, is not very strong. 
This suggest that a small reduction in the local stretching and compression of magnetic field contributes towards the saturation of the fluctuation dynamo.

Before concluding this section, we note that some of these conclusions are similar to those reached independently in the PhD thesis of Denis St-Onge \cite{SO19}, albeit for a different model setup.

\subsection{Local magnetic energy balance}
\label{sec:sat3}
We now directly consider the equation for magnetic energy evolution and calculate its local growth and dissipation terms. 
For an incompressible flow in a periodic domain, the magnetic energy evolution equation can be written as \citep{Roberts67}
\begin{equation}
\frac{d E_M}{d t} = \int_V b_i b_j S_{ij} \, \dd V - \eta \int_V (\nabla \times \vec{b})^2 \, \dd V,
\label{mage}
\end{equation}
where $ E_M = \frac{1}{2} \int_V \vec{b}^2 \, \dd V$ and summation over repeated indices is understood. The term contributing to the energy growth, $b_i b_j S_{ij}$, is calculated at each point in the volume as follows. 
First, we project the magnetic field vector $\vec{b}$ on to each of the eigenvectors of the rate of strain tensor, $\vec{e}_1, \vec{e}_2, \vec{e}_3$.
Let these be $\vec{b}_1, \vec{b}_2, \vec{b}_3$, and then the local growth term $b_i b_j S_{ij} =  \lambda_1 \vec{b}_1^2 +  \lambda_2 \vec{b}_2^2 + \lambda_3 \vec{b}_3^2$ at each position.
This term can be positive or negative ($\lambda_1 < 0$ and $\lambda_3 > 0$). A negative local growth term leads to a decrease in the magnetic energy, whilst a positive value leads to an increase.
The term contributing to the decay in energy is calculated by computing $(\nabla \times \vec{b})^2$ ($\eta=\rm constant$) at each point in space.  

\begin{figure*}
\includegraphics[width=\columnwidth]{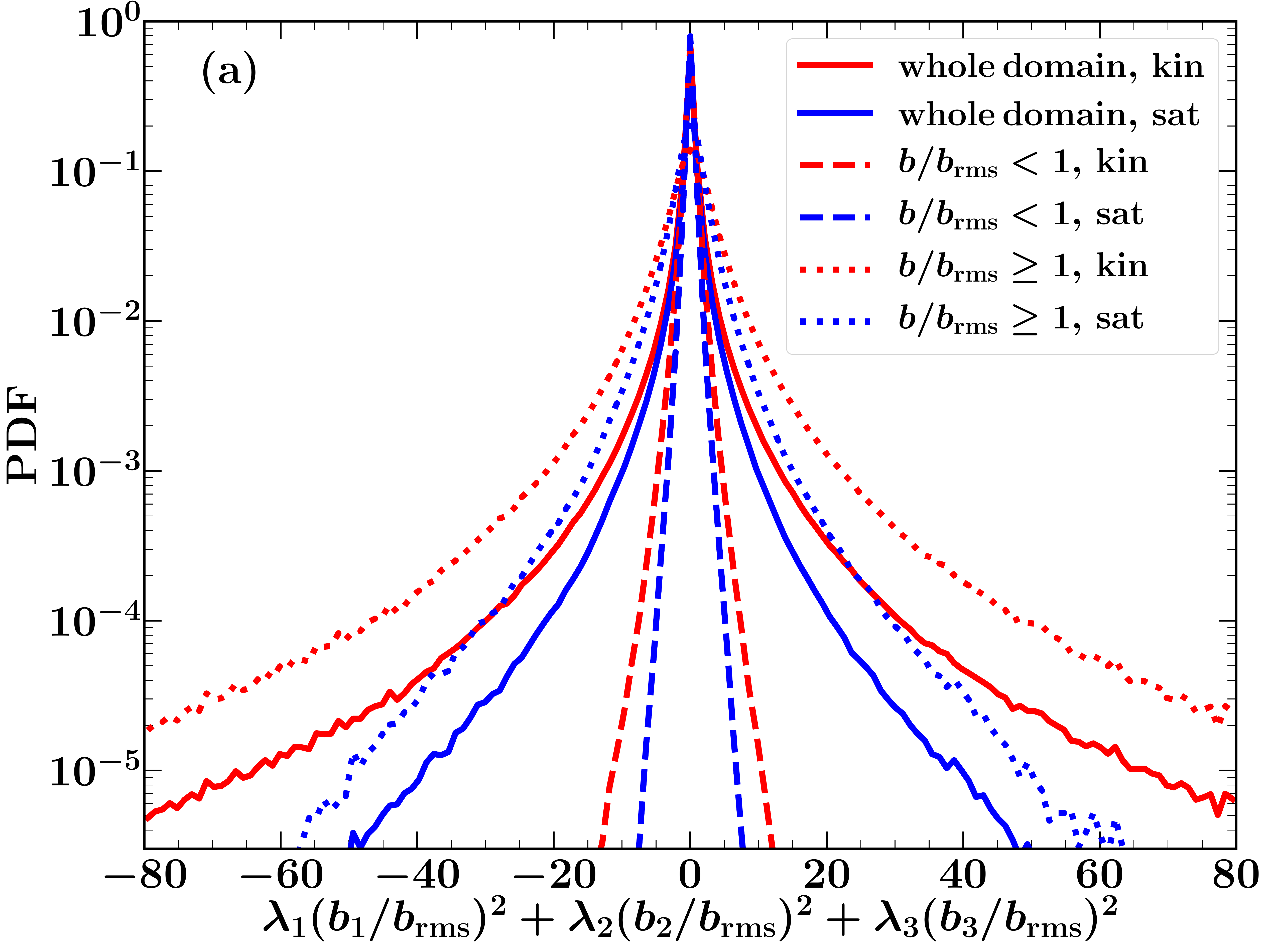}
\includegraphics[width=\columnwidth]{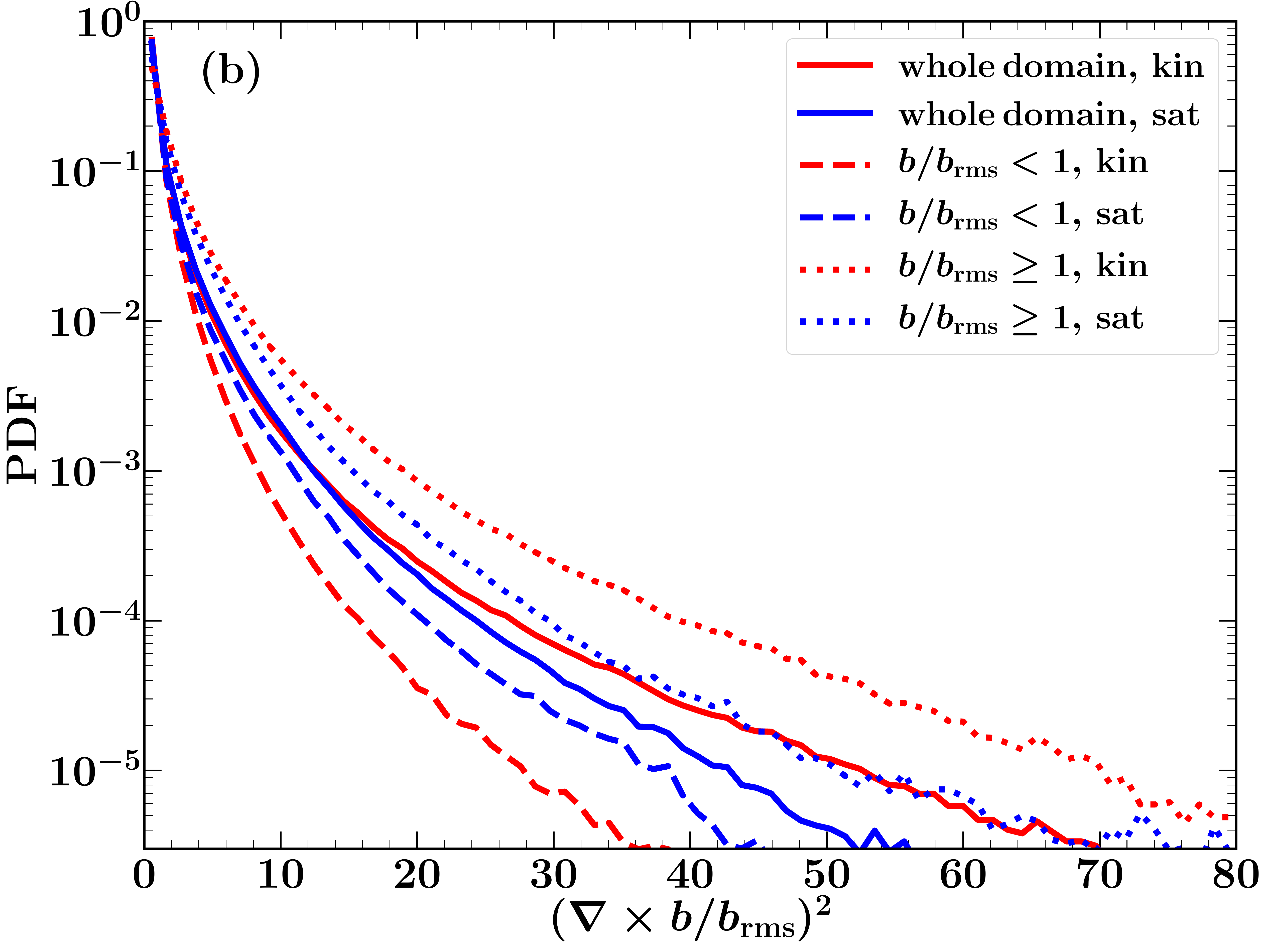}
\caption{The total and conditional PDFs of the local growth term $\lambda_1 (\vec{b}_1/\brms)^2 +  \lambda_2 (\vec{b}_2/\brms)^2 + \lambda_3 (\vec{b}_3/\brms)^2$  (a)
and the local dissipation term $(\nabla \times \vec{b})^2$ (b) in the kinematic (red) and saturated (blue) stages for $\Rm=1122$.
The skewness of the local growth term distribution (solid red line in (a)) is $0.4$ in the kinematic stage and $0.1$ in the saturated stage (solid blue line), so the tendency of this term to promote growth decreases on saturation, as could be expected.
The local dissipation term (b) also decreases statistically as the field saturates.
This conclusions hold in both the weak and strong field regions, except for the local dissipation term, which 
increases in the weak field regions.}
\label{sbj21} 
\end{figure*}
 
\begin{figure*}
\includegraphics[width=\columnwidth]{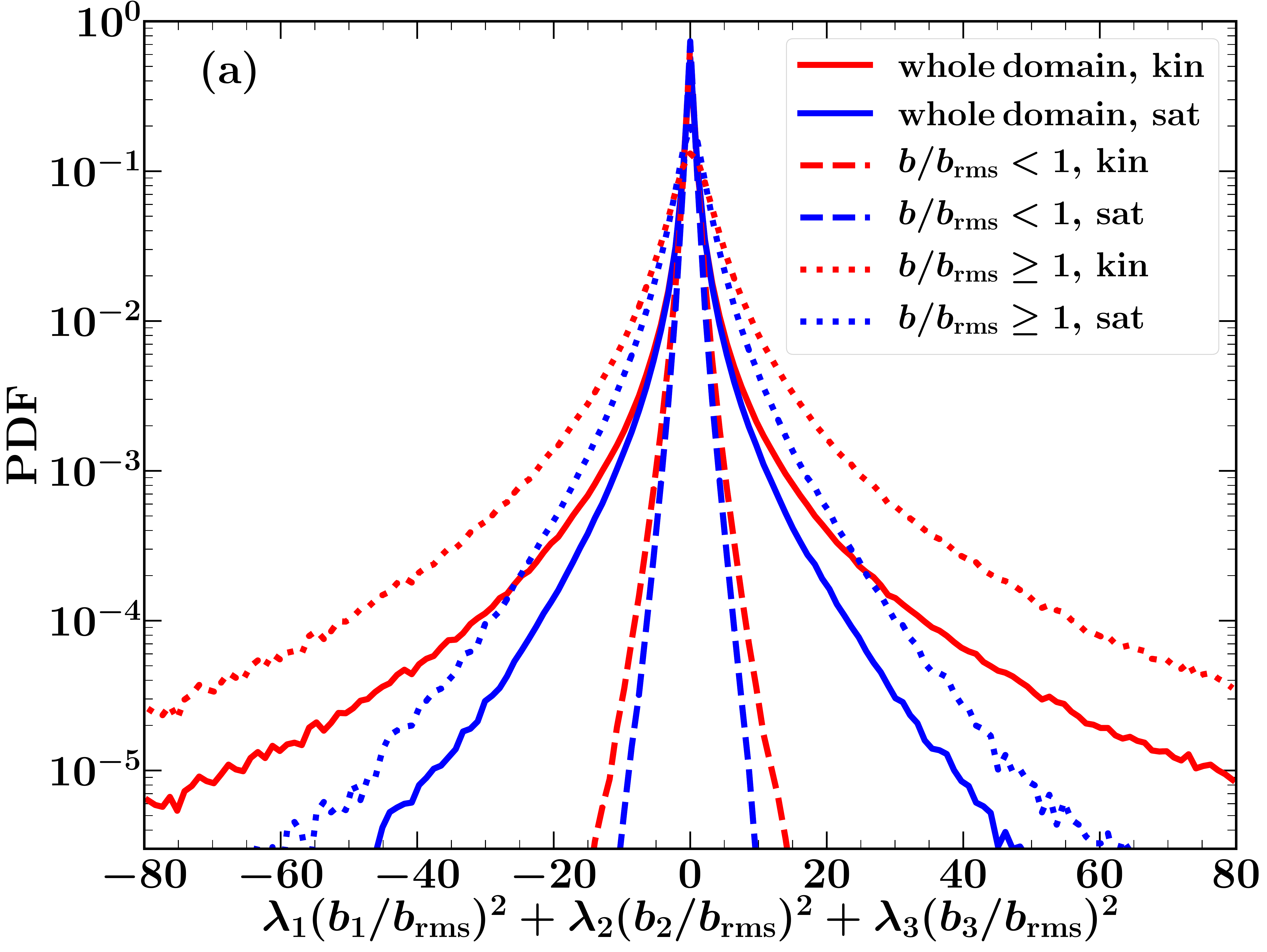}
\includegraphics[width=\columnwidth]{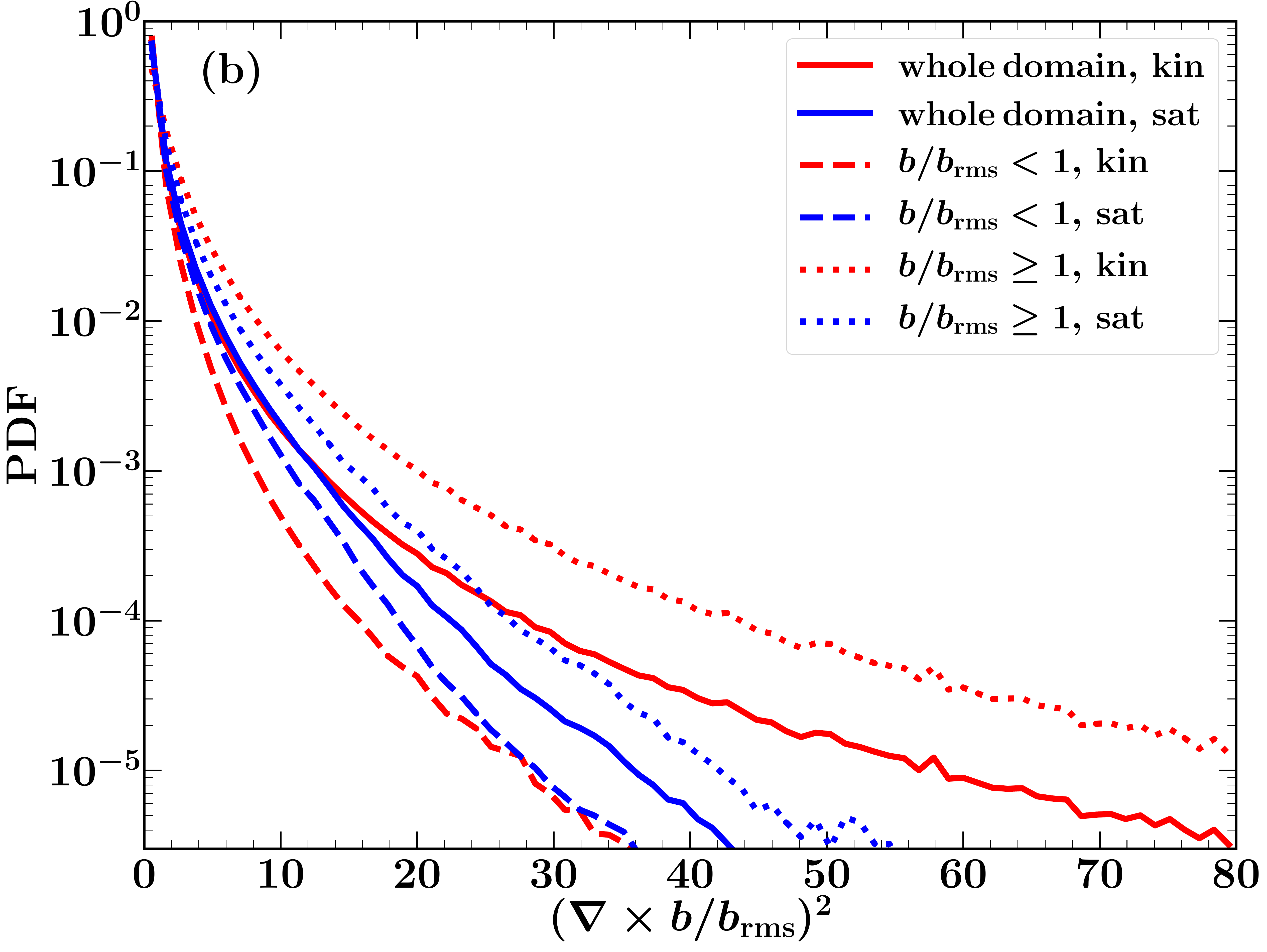}
\caption{As \Fig{sbj21} but for $\Rm=1796$. The skewness of the local growth term distribution (solid red line in (a)) is $0.9$ in the kinematic stage and $0.4$ in the saturated stage (solid blue line).
The conclusions remain the same here as for $\Rm=1122$ in \Fig{sbj21}.}
\label{sbj25} 
\end{figure*}

\Fig{sbj21} and \Fig{sbj25} show the total and conditional PDFs of the local growth and dissipation terms
in the kinematic and saturated stages for $\Rm = 1122$ and $\Rm=1796$ respectively. \Fig{sbj21}a and \Fig{sbj25}a show that the local growth term decreases
on saturation and this is equally true of the strong and weak field regions. This confirms that the stretching of the magnetic field line reduces, which in turn decreases the amplification. 
Numerically, this can be quantified by calculating the skewness of the local growth term distribution in the kinematic and saturated stages (solid red and blue lines in \Fig{sbj21}a and \Fig{sbj25}a).
The skewness is defined for a quantity $X$ as $\langle (X - \langle X \rangle)^3 \rangle / \langle (X -  \langle X \rangle)^2 \rangle^{3/2}$, where $\langle \cdots \rangle$ refers to the mean. The skewness 
of the local growth term distribution in the kinematic (solid red line in \Fig{sbj21}a) and saturated (solid blue line in \Fig{sbj21}a) stage for $\Rm=1122$ are $0.4$ and $0.1$ respectively. The corresponding values
for $\Rm=1796$ (\Fig{sbj25}a) in the kinematic and saturated stages are $0.9$ and $0.4$ respectively. 
The local growth term always has a positive skewness implying continuous magnetic field generation. The skewness decreases on saturation, where the growth is only required to compensate the dissipation. 
The dissipation term also exhibits an overall decrease on saturation as shown in \Fig{sbj21}b and \Fig{sbj25}b, but its behaviour differs in the strong and weak field regions, 
where the dissipation increases in the latter regions.

\begin{figure} 
{\includegraphics[width=\columnwidth]{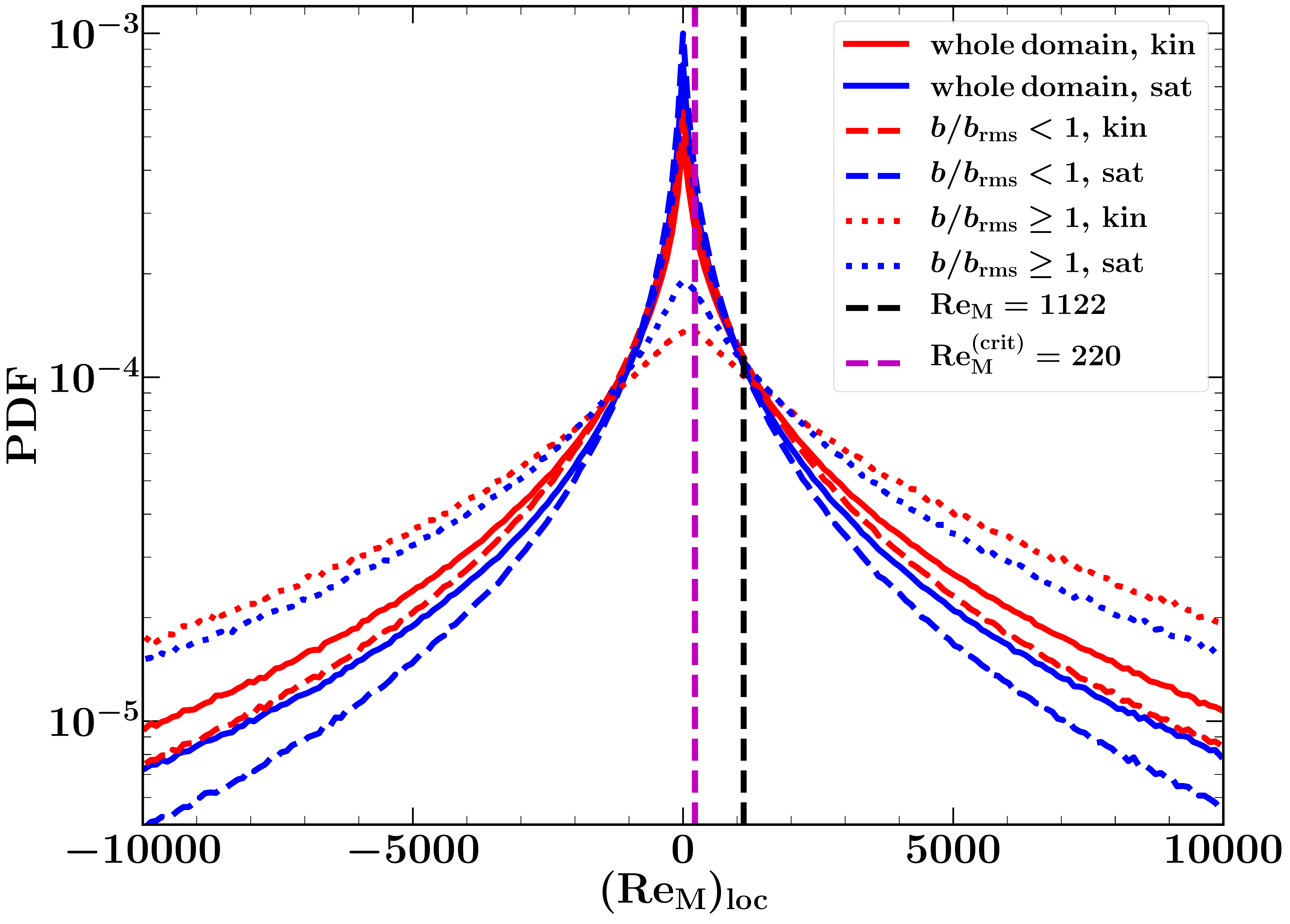}}
\caption{The total and conditional PDFs of the local magnetic Reynolds number $\Rmloc$ in the kinematic (red) and saturated (blue) stages 
with $\Rm=1122$. The purple dashed line shows the critical magnetic Reynolds number $\Rmc=220$ and the black dashed line shows
$\Rm$ for this run.}
\label{rm1} 
\end{figure}

\begin{figure}
  {\includegraphics[width=\columnwidth]{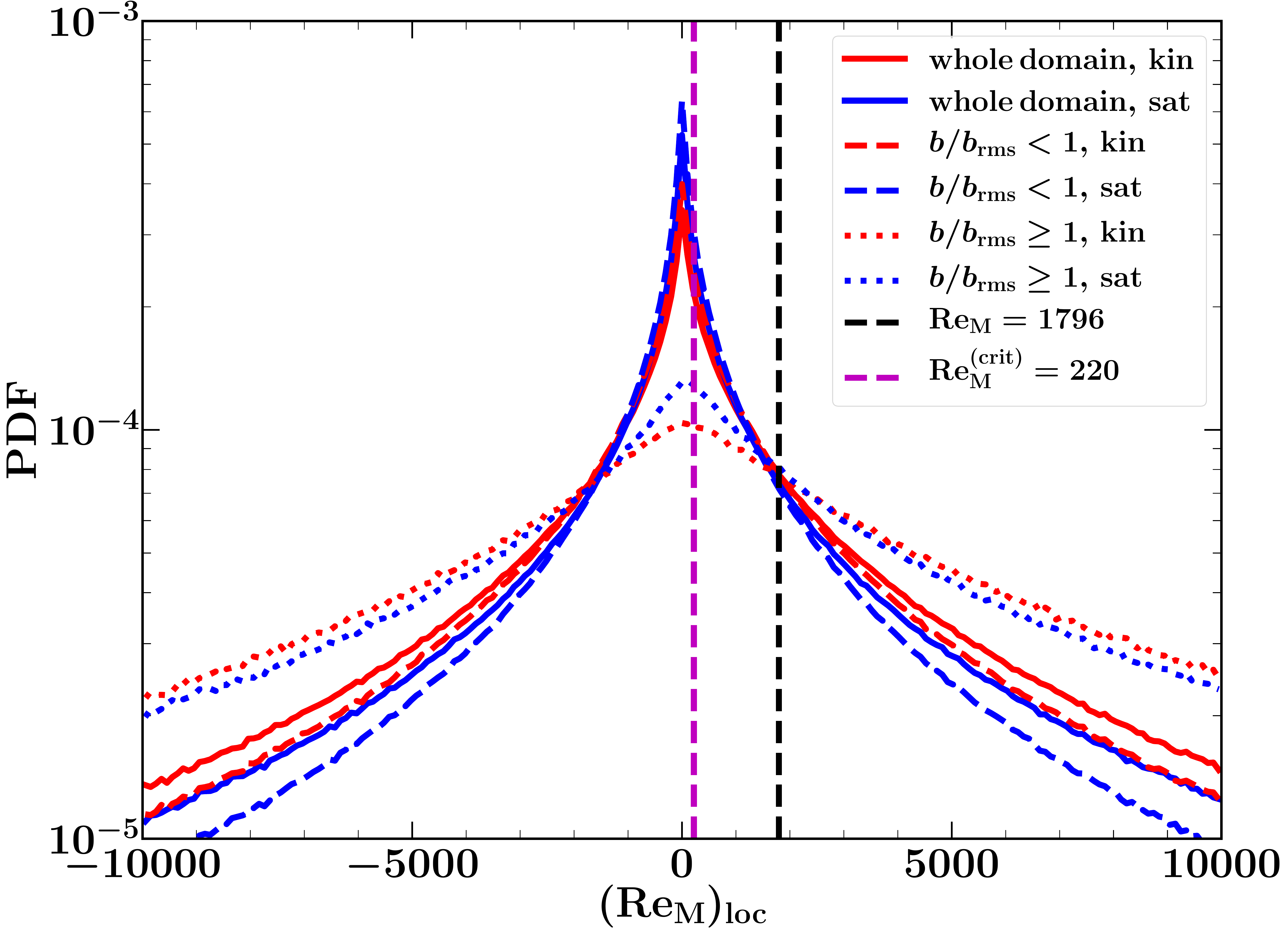}}
\caption{As \Fig{rm1} but for $\Rm=1796$.}
\label{rm25} 
\end{figure}
To calculate the overall decrease or increase in the magnetic energy at each point in the domain, we calculate the local
magnetic Reynolds number. This helps us to explore the behaviour of the diffusion term $\left(\eta \nabla^2 \vec{b}\right)$ in the induction equation (\Eq{fdie}) as the dynamo saturates.
Both terms in \Eq{mage} are calculated at each point in the volume, and the local
magnetic Reynolds number is derived at each position as
\begin{align}
\Rmloc = \frac{b_i b_j S_{ij}}{\eta  (\nabla \times \vec{b})^2},
\label{rmloc1}
\end{align}
providing a measure of the local dynamo efficiency. The local magnetic Reynolds number can be positive or negative, signifying the locally
increasing or decreasing magnetic field strength, respectively. \Fig{rm1} and \Fig{rm25} show the total and conditional PDFs of the local magnetic Reynolds number in the kinematic
and saturated stages for $\Rm=1122$ and $\Rm=1796$. $\Rmloc$ varies from values much less than to those much greater than $\Rmc$ in both the kinematic and 
saturated stages. Thus, magnetic field grows and decays in different parts of the volume but remains in a statistically
steady state overall in the saturated stage. On saturation, both \Fig{rm1} and \Fig{rm25} show that $\Rmloc$
decreases statistically. The mean of $\Rmloc$ for $\Rm=1122$ in the kinematic stage is $808$ and that in the saturated stage is $595$. 
Thus, the mean value of the local magnetic Reynolds number over the entire domain decreases on saturation (to a value close to but not exactly equal to the critical value, $\Rmc\approx220$). 
This effectively implies a relative enhancement in the local diffusion in comparison to the local stretching, which also contributes towards the saturation of the fluctuation dynamo.

To summarize, the fluctuation dynamo saturates due to both reduction in stretching and altered diffusion.
The alignment between the velocity and magnetic fields increases as the field saturates, signifying reduced amplification.
Furthermore, the current density and magnetic field are also statistically better aligned in the saturated stage, which implies a trend towards a force--free field.  
The local growth term statistically decreases (the skewness of the distribution, though remaining positive, decreases on saturation), which implies that the
reduced magnetic field stretching reduces the amplification, which contributes towards the saturation of the fluctuation dynamo.
The local magnetic Reynolds number, though varying over a wide range from values much less than to much higher than the critical value, 
decreases on average. This further implies relative enhancement in the local dissipation compared to the local stretching, which also contributes towards the saturation of the fluctuation dynamo.

\section{Morphology of magnetic structures}
\label{sec:morph}
As shown in \Sec{sec:int}, magnetic field generated by a fluctuation dynamo is intermittent as it is concentrated in filaments, sheets 
and ribbons (\Fig{2db} and \Fig{3db}). To characterize the magnetic structures, 
we use the Minkowski functionals \citep{Min1903}. 
Minkowski functionals have been used in studying morphology of 
structures in a number of numerical simulations \citep{SBMSSS99,Wilkin2007,LSD2012,ZBPT14,KCAP18,Bag2018,Bag2019} and observations \citep{SG98,BSSSY2000,MFS15,Joby2019}.
\begin{table}
        \centering
           \caption{Four Minkowski functionals (MF) $V_0, V_1, V_2$ and $V_3$, their geometrical interpretation and definitions in three dimensions.
        $\dd V$ is the volume element, $\dd S$ is the surface element, and 
        $\kappa_1$ and $\kappa_2$ are the principle curvatures of the 
        surface of a structure.}
        \begin{tabular}{ccc}
        \hline
        MF & Geometric interpretation & Expression \\
        \hline
         $V_0$ & Volume & $\iiint \dd V$ \\
         $V_1$ & Surface area & $(1/6) \iint \dd S$ \\
         $V_2$ & Integral mean curvature & $(1/6 \pi) \iint (\kappa_1 + \kappa_2) \, \dd S$ \\
         $V_3$ & Euler characteristic & $(1/4 \pi) \iint (\kappa_1 \kappa_2) \, \dd S$ \\
        \hline
        \end{tabular}
      
        \label{tab:mink}
\end{table}
The morphology of a 
$d$--dimensional structure can be described by $d+1$ 
Minkowski functionals. In three dimensions, there are four Minkowski 
functionals, as described in \Tab{tab:mink}. We calculate the
Minkowski functionals using Crofton's formulae
\citep{Crofton68, LKM2011} and then calculate the representative length 
scales ($l_1,l_2,l_3$) of magnetic structures (defined by isosurfaces at a fixed value of the magnetic field strength, e.g., see \Fig{3db}) as \citep{SSS98,SBMSSS99}
\begin{equation}
l_1 = \frac{V_0}{2 V_1}, \quad l_2 = \frac{2 V_1}{\pi V_2}, \quad 
l_3 = \frac{3 V_2}{4 V_3}. \label{deftwl} 
\end{equation}
We associate the smallest of these length scales with the thickness $T$ of the structures,
the next largest with the width $W$ and the largest
length scale with the length $L$, i.e., if $l_1 \le l_2 \le l_3$, then
$T=l_1, W=l_2 \, \text{and} \, L=l_3$. The thickness, width and length 
 can be further used to obtain dimensionless measures of the structure shape: planarity $p$ and filamentarity $f$, given by 
\begin{equation}
p = \frac{W-T}{W+T}, \quad f = \frac{L-W}{L+W}.
\label{defpf}
\end{equation}
By definition, 
$0 \le p \le 1$ and $0\le f \le 1$;
$p=0$ and $f=1$ for a perfect filament, $p=0$ and $f=0$ for a sphere, and $p=1$ and $f=0$ for a sheet.
The planarity and filamentarity are not sensitive to the size of the structures but quantify the shape. It is useful to remember that, unlike the Minkowski functionals, $p$ and $f$ are not additive.

\begin{table}
	\centering
	\caption{Parameters of various runs for the nonlinear fluctuation dynamo in a numerical domain size of $(2\pi)^3$ with $512^3$ mesh points.
	In all cases, the forcing scale is approximately $L/5$, the forcing amplitude is $F_0\approx0.02$ and the hydrodynamic viscosity is $\nu=4\times 10^{-4}$.
	The magnetic diffusivity $\eta$, the rms velocity in the saturated stage $\urms$, the Reynolds number $\Re$, 
	the magnetic Reynolds number $\Rm$, the magnetic Prandtl number $\Pm$ and the critical magnetic Reynolds number $\Rmc (\approx 220 \Pm^{-1/2})$ are given.}
	\label{table_nfdkf5}
	\begin{tabular}{cccccc} 
		\hline 
		$\eta$ & $\urms$ & $\Re$ &$\Rm$ & $\Pm$  & $\Rmc$\\
		\hline 
                 $4\times 10^{-4}$ & $0.11$ & $346$ & $346$ & $1.00$ & $220$ \\
                 $3\times 10^{-4}$ & $0.11$ & $346$ & $461$ & $1.33$ & $191$\\
                 $2\times 10^{-4}$ & $0.10$ & $314$ & $628$ & $2.00$ & $156$ \\
                 $1\times 10^{-4}$ & $0.09$ & $283$ & $1131$ & $4.00$ & $110$\\
                 $7.5\times 10^{-5}$ & $0.09$ & $283$ & $1508$ & $5.33$ & $95$\\
                 $5\times 10^{-5}$ & $0.09$ & $283$ & $2261$ & $8.00$ & $78$\\
		\hline
	\end{tabular}
\end{table}
To explore the morphology of magnetic structrures for a range of $\Rm$ values, we use simulations with parameters given in \Tab{table_nfdkf5}. 
We keep $\Re$ about the same for all runs, vary $\Rm$ (making sure $\Pm \ge 1$), and choose $\kf \approx 5 (2 \pi/L)$, so there is a sufficient number of magnetic correlation cells within the volume (with $5^3$ velocity correlation cells). 

\begin{figure*}
\includegraphics[width=\columnwidth]{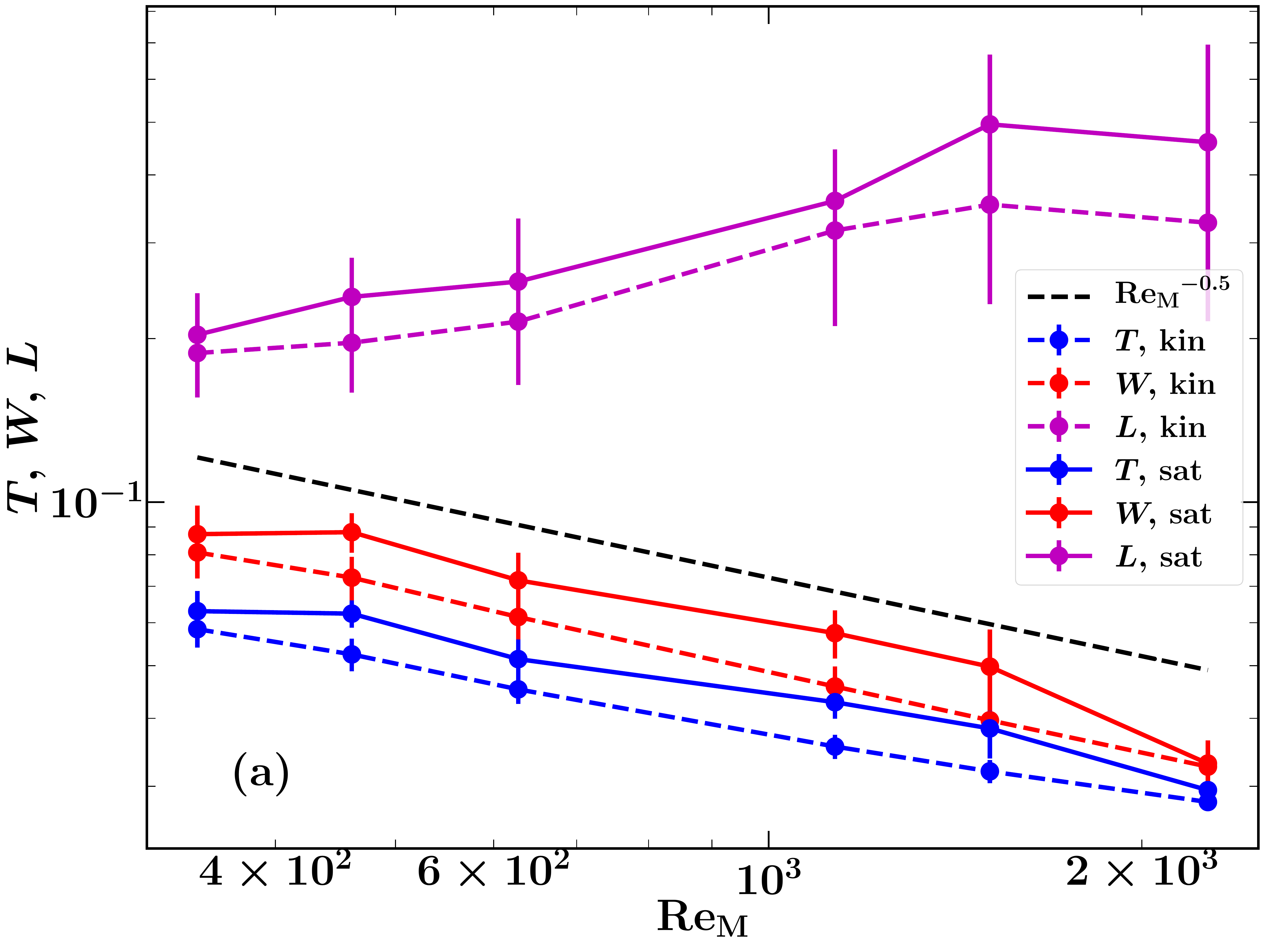}
\includegraphics[width=\columnwidth]{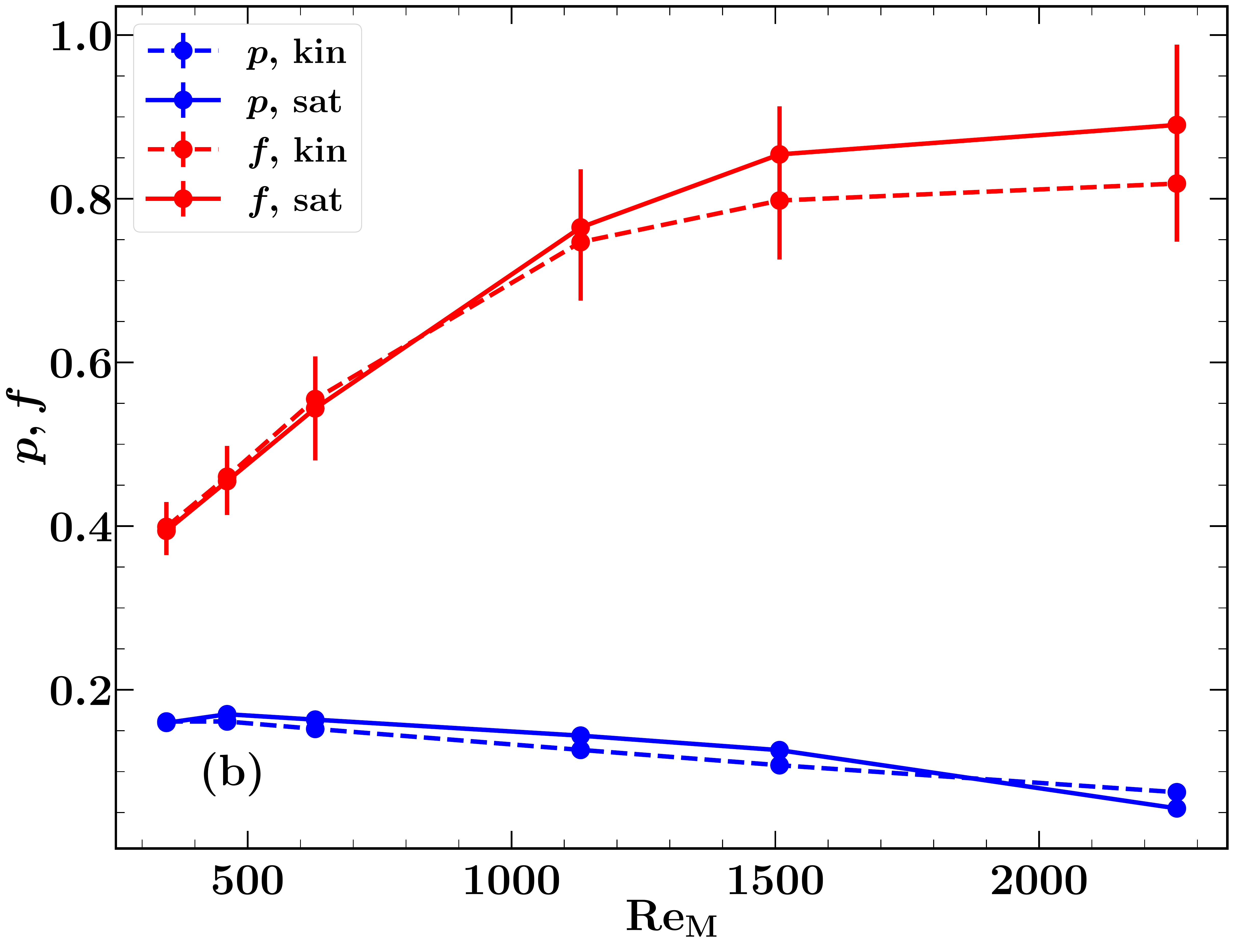}
\caption{(a) Average length ($L$), thickness ($T$) and width ($W$) of magnetic structures in the kinematic (dashed, color) and saturated stages (solid, color) of the nonlinear fluctuation dynamo as functions
of $\Rm$. The width and thickness of the magnetic structures both decrease as $\Rm^{-0.5}$. The $\Rm$ dependence of the size of the structures
is approximately the same in both the kinematic and saturated stages. (b) Planarity ($p$) and filamentarity ($f$) of the magnetic structures, as functions of $\Rm$, for the kinematic (dashed, color) and 
saturated (solid, color) stages. As $\Rm$ increases, the filamentarity increases and the planarity decreases but they seem to approach an asymptotic value after $\Rm\approx1200$. 
The trend with respect to $\Rm$ is the same for both the kinematic and saturated stages.}
\label{mf} 
\end{figure*}

\Fig{mf}a shows the thickness, width and length of magnetic structures obtained by averaging over $30$ values of magnetic field strengths ranging from $b/\brms=2.5$ 
to $4$. The lower limit of the magnetic field strength is chosen to ensure that the structures represent the tail of the PDF (e.g., see \Fig{bmpdf}), whilst the upper limit is chosen to ensure a sufficient
number of points within each structure. The computed values of planarity and filamentarity also remain roughly constant within this selected range of magnetic field strengths.
For the kinematic stage, we expect that the largest length scale $L$ will be independent of $\Rm$. This is because the length of the structures is controlled by the correlation length of the flow 
since the magnetic correlation function of the fastest growing dynamo mode decreases exponentially after that scale \citep{ZRS90}.
As seen in \Fig{mf}a, the length remains roughly constant but then increases slightly after $\Rm\approx600$
and again remains roughly constant. This variation is likely to be due to the decrease in the Reynolds number $\Re$ (\Tab{table_nfdkf5}). 
The other two scales ($W$ and $T$) decrease as $\Rm^{-0.5}$. This scaling can be obtained by balancing the rate of magnetic dissipation with the local shearing rate \citep{Sub98}, 
$\eta/W^2 \simeq \urms/l_0$, where $\eta$ is the magnetic resistivity, $\urms$ is the rms turbulent velocity and $l_0$ is the driving scale
of the turbulence. This gives $W \simeq l_0 (\eta/\urms l_0)^{1/2} = l_0  \Rm^{-0.5}$.
This means that the shape of the magnetic structures becomes more filamentary ($L\gg W\approx T$) and ribbon-like ($T\lesssim W\ll L$) as $\Rm$ increases, but the filamentarity is always larger than the planarity, so the filaments dominate among the magnetic structures
\footnote{For $\Pm > 1$, fast magnetic reconnection at very high $\Rm$ might alter the shape of the structures \citep{Rincon19}, see \Sec{sec:dis} for further discussion.}.
The differences in $\Rm$ scalings with the previous work \citep{Wilkin2007} is probably due to the following reasons. First, they have a prescribed velocity field with forcing at a range of scales, whereas
we force the flow at two scales ($k=4$ and $k=6$) and then let it evolve via the Navier-Stokes equation.
Second, our simulations are at a higher resolution $\left(512^3\right)$ as compared to theirs $\left(128^3\right)$ and thus magnetic structures, especially at higher $\Rm$, are better resolved in our case.
Last and most importantly, they consider values of $\Rm$ which are both lower and higher than $\Rmc$, whereas we only consider $\Rm > \Rmc$. This is because we strongly believe
that those two regimes $\left(\Rm < \Rmc \, \text{and} \, \Rm \ge \Rmc \right)$ are physically different and must not be considered together to characterize the length scales of magnetic structures as functions of $\Rm$.

All three scales are larger in the saturated stage than in the kinematic stage. 
Thus, the magnetic structures become larger as the magnetic field saturates. This is also the reason that the magnetic field correlation length scale increases as the field saturates (as shown in \Tab{table_nfd}).
The increase in the length (the largest length scale) of magnetic structures on saturation is consistent with the finding by Schekochihin et al.  \citep{SCTMM04}.
The $\Rm$ scaling for all three scales is roughly the same for both the kinematic and saturated stages. 

\Fig{mf}b shows the planarity and filamentarity of magnetic structures as functions of $\Rm$.
The filamentarity is always higher than the planarity and thus the magnetic structures are more like filaments in both the kinematic and saturated stages.  
The dependence of these morphological measures on $\Rm$ is the same for the kinematic and saturated stages.

\section{Conclusions and discussion} 
\label{sec:dis}
It is important to understand the saturated state of the fluctuation dynamo because the saturated state seeds the mean field dynamo, controls the small-scale magnetic field
structure and decides the magnetic field length scales in the system where the mean field dynamo is absent (for example, elliptical galaxies). Moreover, it is crucial to understand
the physics of the saturation mechanism because numerical simulations, at present, are at much lower values of $\Rm$ than their estimated values ($\Rm \approx 10^{18}$
for spiral galaxies, $\Rm \approx 10^{22}$ for elliptical galaxies and $\Rm \approx 10^{29}$ for galaxy clusters). 

Using numerical simulations of driven nearly incompressible turbulence, we have 
explored the saturation mechanism of the fluctuation dynamo. We find that the dynamo saturates because both the amplification and diffusion are affected by the action of the Lorentz force on the flow.
Most previously suggested mechanisms hinted at changes in either of those two and thus required significant changes in the properties of the velocity and magnetic fields 
from the kinematic stage. For example, if only the enhancement in diffusion
is responsible, it would require the effective $\Rm$ in the saturated state to reduce from hugely supercritical levels to values close to $\Rmc$ ($\simeq 10^{2} \text{--} 10^{3}$ \citep{Kazantsev1968}). 
And, if only the decrease in amplification is responsible for saturating the dynamo, it would require a drastic decrease in the Lyapunov exponents (which are a measure of chaotic properties of the flow) \citep{CHK96}. We
suggest that both occur and thus such a dramatic change is not necessary. We confirm that the amplification decreases by reduction in the stretching of magnetic field lines. 
The local magnetic Reynolds number $\Rmloc$, which is suggested as a measure of the local magnetic diffusion, decreases slightly. This confirms that the local diffusion of magnetic field relative to field line stretching is enhanced, 
which is also responsible for saturating the dynamo.

The fluctuation dynamo-generated magnetic field is spatially intermittent. So, we studied the morphology of the magnetic structures in the kinematic and saturated stages.
In both cases, the largest length scale is roughly independent of $\Rm$ and the other two scales decrease as $\Rm^{-0.5}$. 
We find that the structures are of a larger size (all three length scales increase) in the saturated stage as compared to the kinematic stage. This agrees with the
results in \Tab{table_nfd}, where we find that the correlation length is higher for the saturated magnetic field. 
This also aligns with the conclusion in the \Sec{sec:int} (also shown in \citep{SCTMM04}) that the magnetic field is less intermittent in the saturated stage as compared to the kinematic stage.
However, the $\Rm$ dependence is the same for both the stages and thus 
the overall shape of magnetic structures produced by the fluctuation dynamo is not affected by the Lorentz force to any significant extent (all three length scales increase but in a very similar way).

The study explores physical effects over a range of $\Rm$ for $\Pm \ge 1$. 
However, for $\Pm > 1$ at very high $\Rm$ ($\gtrsim 10^3$), the fields might be unstable to fast magnetic reconnection \citep{Rincon19}.
This might change the morphology of magnetic fields, locally affect velocity fields and thus might alter the saturated state of the fluctuation dynamo. However, the effect of fast, stochastic magnetic reconnection
on the dynamo is not very well understood yet \citep{Eyink11} and would require high-resolution numerical simulations over a number of very 
high $\Rm$ values to study the effect of fast magnetic reconnection on the fluctuation dynamo saturation mechanism.

The study can be extended in several ways. An immediate extension would be to repeat 
the entire analysis for dynamos in a stratified medium \citep{HBM04,Fed11,Fed14,SBS18}, which is more relevant for young galaxies and star-forming gas clouds. 
We have performed the analysis for $\Pm \ge 1$ which is of relevance to fluctuation dynamo in the interstellar and intergalactic medium but this should be extended to
the $\Pm < 1$ regime which is important for stars, planets and liquid metal experiments \citep{Brandenburg11,Sahoo11}.
We have adopted the MHD approximation but plasma effects might also play an important role. It would also be interesting to compare our results with those of the plasma dynamo \citep{Rincon16,DK18} and see
how the relationship between velocity and magnetic fields and the magnetic field structure change when plasma effects are considered. Plasma effects might be particularly
important for the weakly collisional gas in galaxy clusters. We aim to consider such problems in our future work.
\begin{acknowledgments} 
We thank Kandaswamy Subramanian and Christoph Federrath for useful discussions and comments on the paper. We acknowledge financial support of the STFC (ST/N000900/1, Project 2) 
and the Leverhulme Trust (RPG-2014-427). We thank one of the referees for highlighting Denis St-Onge's PhD thesis \citep{SO19}, which independently reports results similar to those found in \Sec{sec:sat2}. In fact, \Fig{lrun25} was added following the initial review so as to facilitate a direct comparison between the two studies.
\end{acknowledgments}
%
\end{document}